\documentclass[10pt,twocolumn,letterpaper]{article}

\usepackage[pagenumbers]{iccv} 
\usepackage{times}
\usepackage{epsfig}
\usepackage{graphicx}
\usepackage{subcaption}
\usepackage{natbib}
\usepackage{booktabs}
\usepackage{multirow}
\usepackage{amsmath}
\usepackage{amssymb}
\usepackage{pifont}
\usepackage{newfloat}
\usepackage{listings}
\usepackage{caption}
\usepackage{ulem}
\usepackage{colortbl}
\usepackage{xcolor}
\usepackage{float}
\usepackage{afterpage}
\definecolor{iccvblue}{rgb}{0.21,0.49,0.74}
\usepackage[pagebackref,breaklinks,colorlinks,allcolors=iccvblue]{hyperref}

\def\paperID{9793} 
\def\confName{ICCV}
\def\confYear{2025}

\title{FM2S: Towards Spatially-Correlated Noise Modeling\\in Zero-Shot Fluorescence Microscopy Image Denoising}

\author{
Jizhihui Liu\textsuperscript{1}, Qixun Teng\textsuperscript{1}, Qing Ma\textsuperscript{2}, 
Junjun Jiang\textsuperscript{1}\thanks{Corresponding author: Junjun Jiang} \\  
\textsuperscript{1}Harbin Institute of Technology \\  
\textsuperscript{2}The Hong Kong Polytechnic University \\  
{\tt \small \{danielement321, choloteteng, qingma2016\}@gmail.com} \\
{\tt \small jiangjunjun@hit.edu.cn}
}

\begin{document}
\maketitle
\begin{abstract}
   Fluorescence microscopy image (FMI) denoising faces critical challenges due to the compound mixed Poisson-Gaussian noise with strong spatial correlation and the impracticality of acquiring paired noisy/clean data in dynamic biomedical scenarios. While supervised methods trained on synthetic noise (e.g., Gaussian/Poisson) suffer from out-of-distribution generalization issues, existing self-supervised approaches degrade under real FMI noise due to oversimplified noise assumptions and computationally intensive deep architectures. In this paper, we propose \textbf{F}luorescence \textbf{M}icrograph to \textbf{S}elf (FM2S), a zero-shot denoiser that achieves efficient FMI denoising through three key innovations: 1) A noise injection module that ensures training data sufficiency through adaptive Poisson-Gaussian synthesis while preserving spatial correlation and global statistics of FMI noise for robust model generalization; 2) A two-stage progressive learning strategy that first recovers structural priors via pre-denoised targets then refines high-frequency details through noise distribution alignment; 3) An ultra-lightweight network (3.5k parameters) enabling rapid convergence with 270× faster training and inference than SOTAs. Extensive experiments across FMI datasets demonstrate FM2S's superiority: It outperforms CVF-SID by 1.4dB PSNR on average while requiring 0.1\% parameters of AP-BSN. Notably, FM2S maintains stable performance across varying noise levels, proving its practicality for microscopy platforms with diverse sensor characteristics. Code and datasets will be released.
\end{abstract}    
\section{Introduction}
\label{sec:intro}
Fluorescence microscopy image (FMI) denoising remains a persistent challenge. 
For many reasons, it lacks sufficient labelled noisy/clean image pairs for training. Capturing the ground truth of Fluorescence microscopy image (FMI) requires long exposure or multiple shots in dynamic scenes with motion, however, which is usually accompanied by photodamage or photobleaching \cite{mannam2022real}. Besides, obtaining a large number of noisy FMI in practice is challenging and sometimes unavailable since it requires multiple shots under the same static scene with several constraints \cite{neshatavar2022cvf}.

Moreover, the noise in FMI is usually much more intense than other common images due to the minimal number of photons the detector can capture \cite{zhang2019poisson} and different in-camera processing pipelines \cite{zhou2020w2s}. In general, the noise consists of quantum noise (correlated with pixel intensity) and thermal noise (independent of pixel intensity) \cite{maji2019feature,mannam2022real}. These can correspond to Poisson and Gaussian noise, respectively, and the FMI noise can be seen as compound Mixed Poisson-Gaussian (MPG) noise \cite{zhang2019poisson}, which is neither additive nor multiplicative noise, limiting most general image denoising methods.

\begin{figure}[tbp]
    \centering
    \begin{subfigure}[b]{0.23\textwidth}
        \centering
        \includegraphics[width=\textwidth]{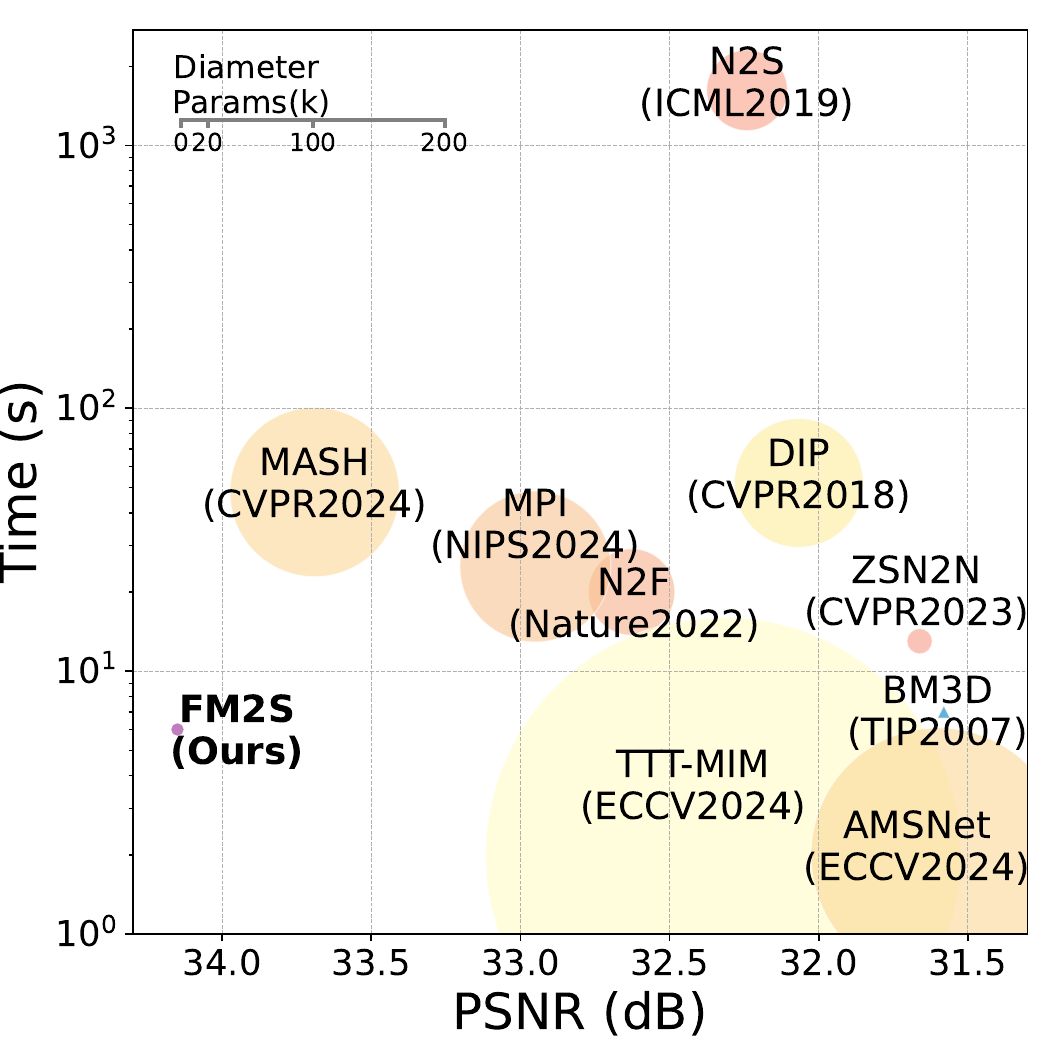}
    \end{subfigure}
    \hfill
    \begin{subfigure}[b]{0.225\textwidth}
        \centering
        \includegraphics[width=\textwidth]{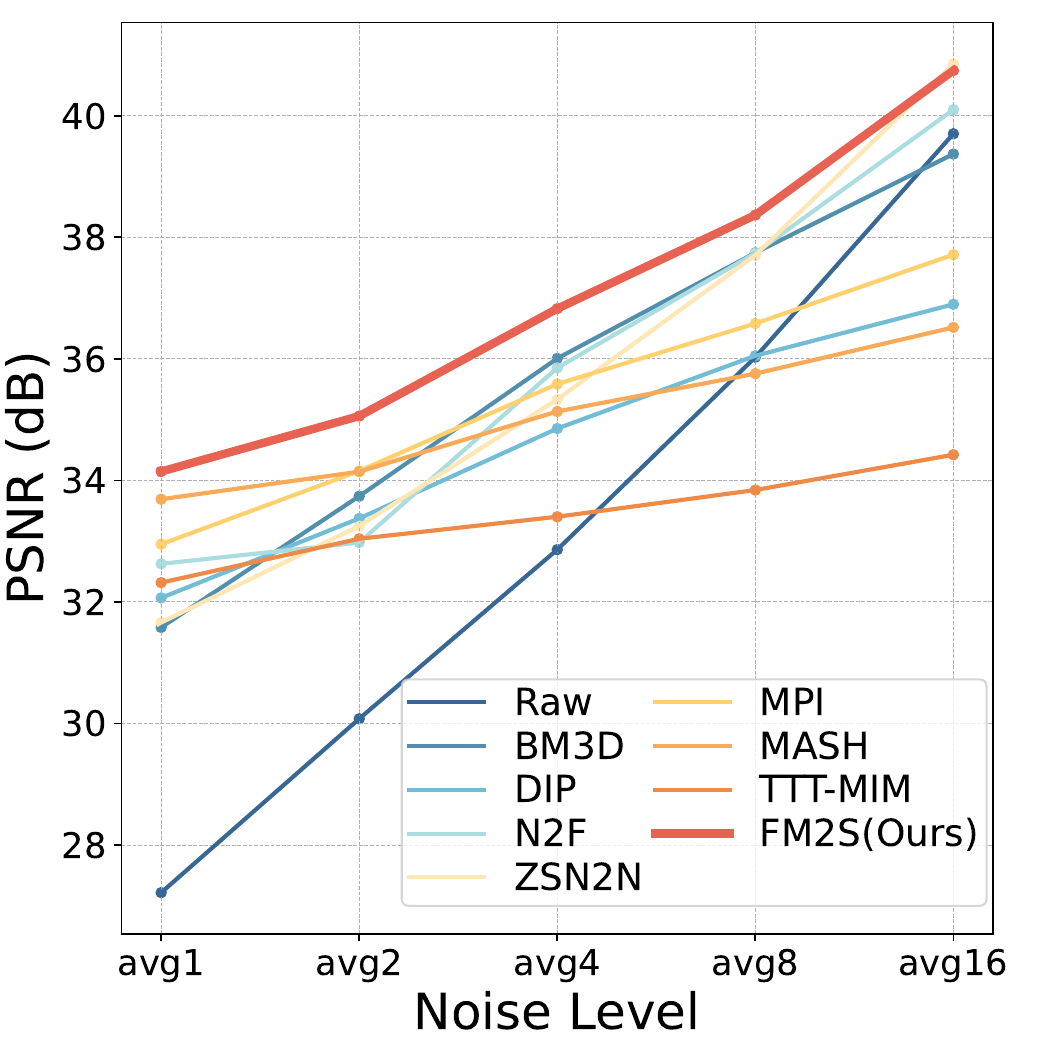}
    \end{subfigure}
    \caption{(a) We propose FM2S for efficient fluorescence microscopy denoising, which features performance and efficiency. The circle size denotes the size of the denoising network for deep-learning-based methods. Note that the y-axis is in the log scale. (b) PSNR \textit{v.s.} noise levels of different denoising methods. The higher avg$N$ stands for a lower noise level.}
    \label{fig:Intro}
\end{figure}

Supervised image denoising methods \cite{zamir2022restormer,cheng2024transfer,neshatavar2022cvf,chen2024exploring,lee2022ap,wang2023lg}
excel in common image denoising, utilizing denoising neural networks (DNN) and large-scale datasets (e.g., SIDD \cite{Abdelhamed_2018_CVPR}). Unfortunately, these denoisers usually result in poor generalization to out-of-distribution (OOD) noise (i.e., different noise levels and noise types) due to the overfitting to the noise present in the training data \cite{Chen_2023_CVPR} and our experiment in Table \ref{table:MainResults} also confirms this. 

In contrast, methods based on few or zero-shot training samples, focusing on only one image or a small number of images with consistent noise types (from the same camera), show promise for
denoising FMI \cite{dabov2007image,ulyanov2018deep,quan2020self2self,mansour2023zero,chihaoui2024masked,Pang_2021_CVPR,ma2024masked}. However, most of them have insufficient performance when dealing with intricate noise patterns in FMI because of the limited training data. To mitigate this, researchers employ data augmentation techniques, such as downsampling or masking, which result in inevitable information loss \cite{Zou_2023_ICCV}. Additionally, the approaches above often rely on a very deep neural network that requires abundant optimization steps and consumes a long time to denoise \cite{quan2020self2self,batson2019noise2self}. At the same time, they also inevitably suffer from overfitting issues due to limited samples and a large number of model parameters to be optimized.

To address generalization collapse under OOD noise and insufficient samples,
we propose \textbf{F}luorescence \textbf{M}icrograph to \textbf{S}elf (FM2S), an efficient zero-shot denoising method specifically designed for FMI. \textit{Given the complexity of noise, FM2S focuses on learning the main feature of the noise rather than the original noise}. Based on our empirical observations of real FMI images, we introduce a Noise Injection module for data augmentation, which captures both the strong spatial correlation and global properties of real noise. To enhance the learning of noise features, the input data is processed by the Pre-Denoise (e.g., Median Filter) to avoid the interference of the original noise, which simultaneously serves as both training target and Noise Injection input. To further facilitate feature learning, we introduce a two-stage training strategy that enables DNN to learn different levels of features in a coarse-to-fine manner. Along with a simple 3.5k-parameter DNN, it takes much fewer steps for optimization. As Fig. \ref{fig:Intro} suggests, FM2S achieves promising results in terms of denoising performance and time consumption across various noise levels.

In summary, our contributions are as follows: 
\begin{itemize}
    \item We carefully investigate the noise properties in FMI and propose an efficient zero-shot denoiser FM2S, which is suitable for practical use.
    \item We specifically design a Noise Injection module for FMI and a 2-stage training strategy, which assist FM2S in conquering the prevalent OOD problem.
    \item FM2S achieves better results than the SOTA $270 \times$ faster and continues to demonstrate outstanding performance across variable noise levels.
    
    
\end{itemize}

\section{Related Works}
\subsection{Training on Paired Noisy-Clean Images}
Supervised methods rely on a large quantity of clean/noisy image pairs, which train denoising deep neural networks on many noisy/truth image pairs, e.g. \cite{zhang2017beyond,anwar2019real,zamir2020learning,quan2021image}. DnCNN \cite{zhang2017beyond} is one of the benchmarking deep image denoisers, which uses a deep CNN for residual learning. Restormer \cite{zamir2022restormer} employs transformer blocks \cite{vaswani2017attention}, while TransCLIP \cite{cheng2024transfer} uses contrastive language-image pre-training model to encode the image information. However, these methods suffer from train-test noise distribution shifts and struggle with unseen noise types/levels under real-world conditions.
Therefore, gathering well-aligned noisy-clean pairs from real-world scenes is both challenging and impractical for FMI, thus constraining such methods in practice.

\subsection{Training on Only Noisy Images}
To overcome the reliance on clean/noisy image pairs, a branch of self-supervised methods is trained with a large number of noisy images only. Among them, Noise2Noise \cite{lehtinen2018noise2noise} demonstrates that denoisers can be trained without clean images, supervised on pairs of independent noisy images of the same scene. Most further advancements \cite{huang2021neighbor2neighbor,krull2019noise2void,wang2022blind2unblind,xie2020noise2same} focus on eliminating the need for noisy image pairs, leveraging the latent information in the image for denoising.  Recently, AP-BSN \cite{lee2022ap} utilized two pixel-shuffle downsamplings to address structured noise. CVF-SID \cite{neshatavar2022cvf} separates latent image and noise through self-supervised losses.
Despite these advances, obtaining a large number of FMI samples is consuming in both time and finance, and their generalization ability to FMI remains a significant challenge.

\subsection{Training on One Single Noisy Image}
Traditional methods NLM \cite{buades2005non} and BM3D \cite{dabov2007image} can denoise a single image but rely on hand-crafted priors, which are constrained by their inability to adapt to complex or unseen noise patterns. 
More recent methods address these limitations by leveraging blind-spot network (BSN). For example, Noise2Self \cite{batson2019noise2self} avoids identity mapping in BSN without paired images; Self2Self \cite{quan2020self2self} integrates dropout with Bernoulli-sampled training; MASH \cite{chihaoui2024masked} adapts BSN masking via noise correlation and introduces a shuffling technique. 
However, BSN-based architectures will lead to information loss at the blind-spot pixels \cite{wang2022blind2unblind}. 
Another type of approach, such as Zero-Shot Noise2Noise \cite{mansour2023zero}, employs downsampling for data augmentation and directly learns augmented sample relationships, but reduces sampling density and loses high-frequency details. 
Furthermore, methods like NAC \cite{xu2020noisy}, Noisier2Noise \cite{Moran_2020_CVPR}, and R2R \cite{Pang_2021_CVPR} try to generate noisy pairs from a single image. These methods require only one image for both training and inference and feature high applicability in FMI denoising.
\section{Proposed Method}

\begin{figure*}[t]
    \centering
    \includegraphics[width=\textwidth]{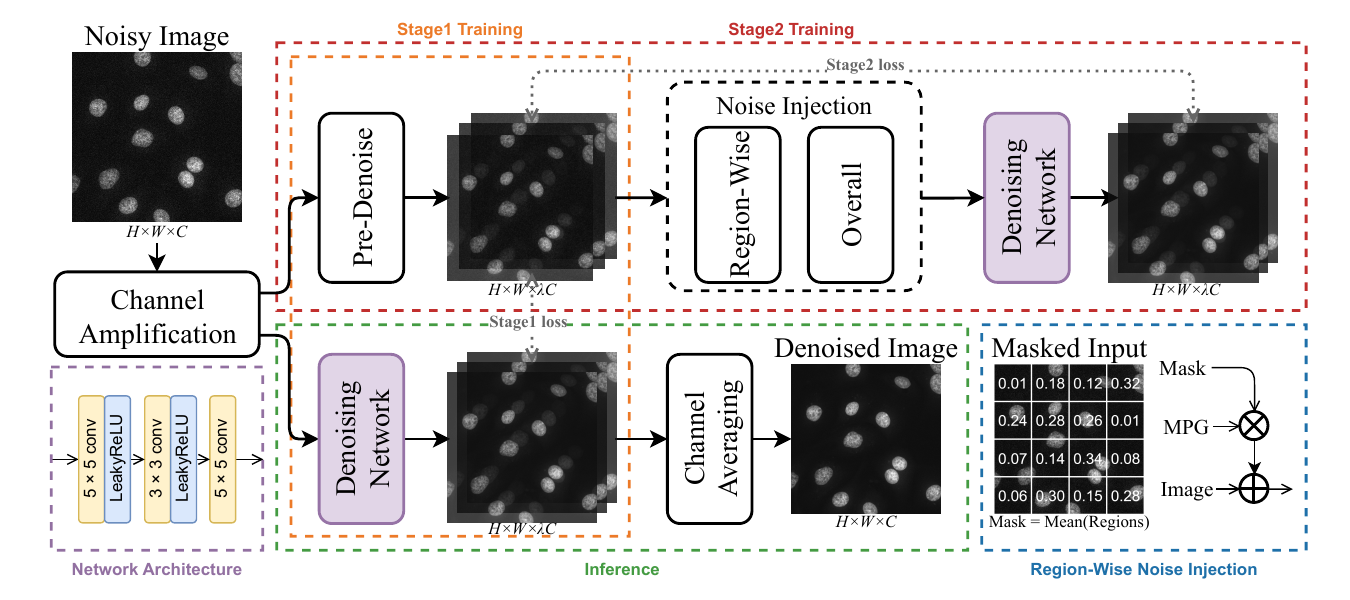}
    \caption{The whole pipeline and key components of FM2S. The original image is first pre-denoised to generate an image with relatively less noise, which will be used for Noise Injection and act as the training target for DNNs. Noise Injection applies adaptive MPG to the filtered image and generates samples for training, where channel amplification is taken to enhance the generalization ability of DNN.}
    \label{fig:Overall}
\end{figure*}

Our main idea is to generate multiple pairs of approximated noisy images and corresponding clean images from a single noisy image and then train a denoising network on these pairs. The whole pipeline is illustrated in Fig. \ref{fig:Overall}. We start by presenting our preliminaries and motivation.

\subsection{Preliminaries}
\label{sec:Preliminaries}
Considering a noisy image $y$ with latent ground truth $x$ in the formulation of
\begin{equation}
    y = \eta (x)
    \label{eq:noiseModel},
\end{equation}
where $\eta ( \cdot )$ denotes noise corruption, which is mostly irreversible. Denoising aims to dig out $x$ with $y$. In a supervised manner\cite{zhang2017beyond, zamir2022restormer, cheng2024transfer}, where abundant noisy images ${y_i},i \in \{ 1, \cdots ,N\} $ and corresponding ground truths ${x_i},i \in \{ 1, \cdots ,N\}$ are given, the network training is to solve
\begin{equation}
    \mathop {\arg }\limits_\theta  \min \sum\limits_i {{{\left\| {{f_\theta }(\eta ({x_i})) - {x_i}} \right\|}_F}},
    \label{eq:optimize with GT}
\end{equation}
where ${{f_\theta }}$ denotes a denoising network with parameter $\theta$ and ${\left\|  \cdot  \right\|_F}$ denotes Frobenius norm. 

However, as mentioned by Section \ref{sec:intro}, pairs of noisy images with ground truths are mostly unavailable in FMI denoising. To mitigate this issue, we generate such image pairs by approximation. The ground truth can be approximated by pre-denoising the noisy image with an elementary denoising method like median filter, BM3D \cite{dabov2007image}, NLM \cite{buades2005non}, etc. For noise approximation, we use a Noise Injection module to generate a large number of noisy samples under the same scene.

Therefore, the basic idea of our proposed method is to transform traditional supervised denoising (Eq. \ref{eq:optimize with GT}) to
\begin{equation}
    \mathop {\arg }\limits_\theta  \min \sum\limits_i {{{\left\| {{f_\theta }(\hat \eta ({u_i})) - {u_i}} \right\|}_F}},
    \label{eq:optimize with filtered}
\end{equation}
where $u$ denotes pre-denoised image and $\hat \eta ( \cdot )$ denotes simulated noise corruption process.

\subsection{Motivation and Observation}
\label{sec:Motivation&Observation}
\noindent \textbf{DCD-Net.} DCD-Net \cite{Zou_2023_ICCV} consists of multiple runs of noise corruption and denoising. The noisy image $y$ is first pre-denoised with a denoise network ${f_\theta }( \cdot )$ to obtain $\hat{x}$. The corruption phase assumes the noise can be modeled by a Gaussian distribution and applies noise to $\hat{x}$ for the construction of noisy pairs $y_1, y_2$ to train ${f_\theta }( \cdot )$ in an N2N manner \cite{lehtinen2018noise2noise}.

The adding-noise scheme in DCD-Net avoids the ignorance of information by blind-spot network and downsampling \cite{Zou_2023_ICCV}. However, the shortcoming is obvious: the Gaussian noise assumption may be insufficient to model the intricate MPG noise in FMI. Besides, DCD-Net relies on a pre-trained N2N network for pre-denoise, which may cause OOD problems in FMI denoising.

\begin{figure}[t]
    \centering
    \includegraphics[width=1\linewidth]{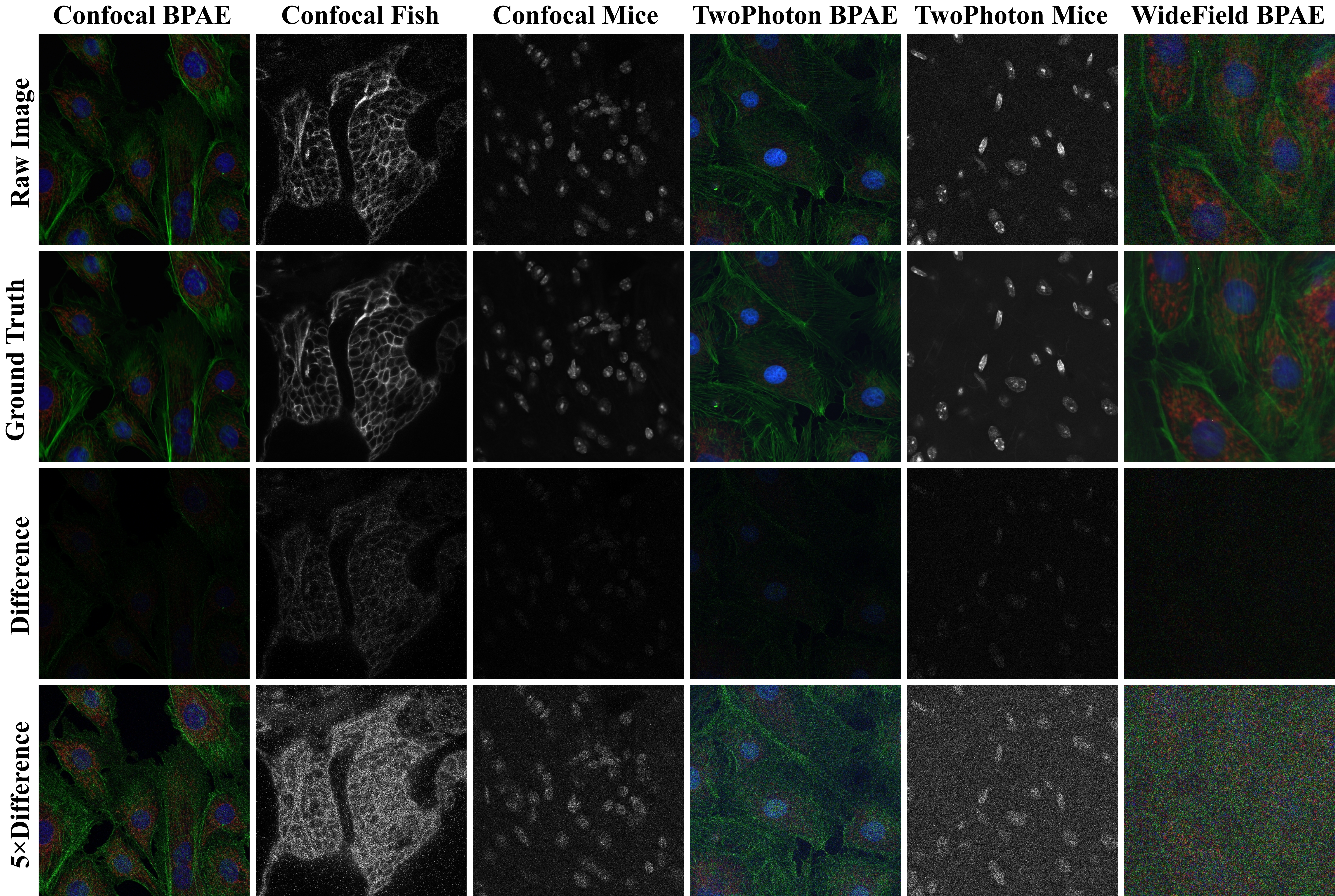}
    \caption{The difference between noisy images and ground truth. The fourth row is $5\times$ brightened difference for better visualization. More examples are in the supplementary material.}
    \label{fig:NoiseObservation}
\end{figure}

\noindent \textbf{Empirical Observation.} To overcome the above problems, having an insight into the FMI noise is crucial. Fig. \ref{fig:NoiseObservation} depicts the gap between noisy images and related ground truths, which is to be minimized. Most of the corruption is in and surrounding the brighter areas, which reflects the signal-dependent Poisson noise, while the corruption diffuses in the whole field, including darker areas, reflecting the signal-independent Gaussian noise \cite{mannam2022real}. This observation is the foundation of the Noise Injection module in Section \ref{sec:NoiseInjection}.

\subsection{Noise Injection Pipeline}
\label{sec:NoiseInjection}
\begin{figure}[htbp]
    \centering
    \includegraphics[width=0.9\linewidth]{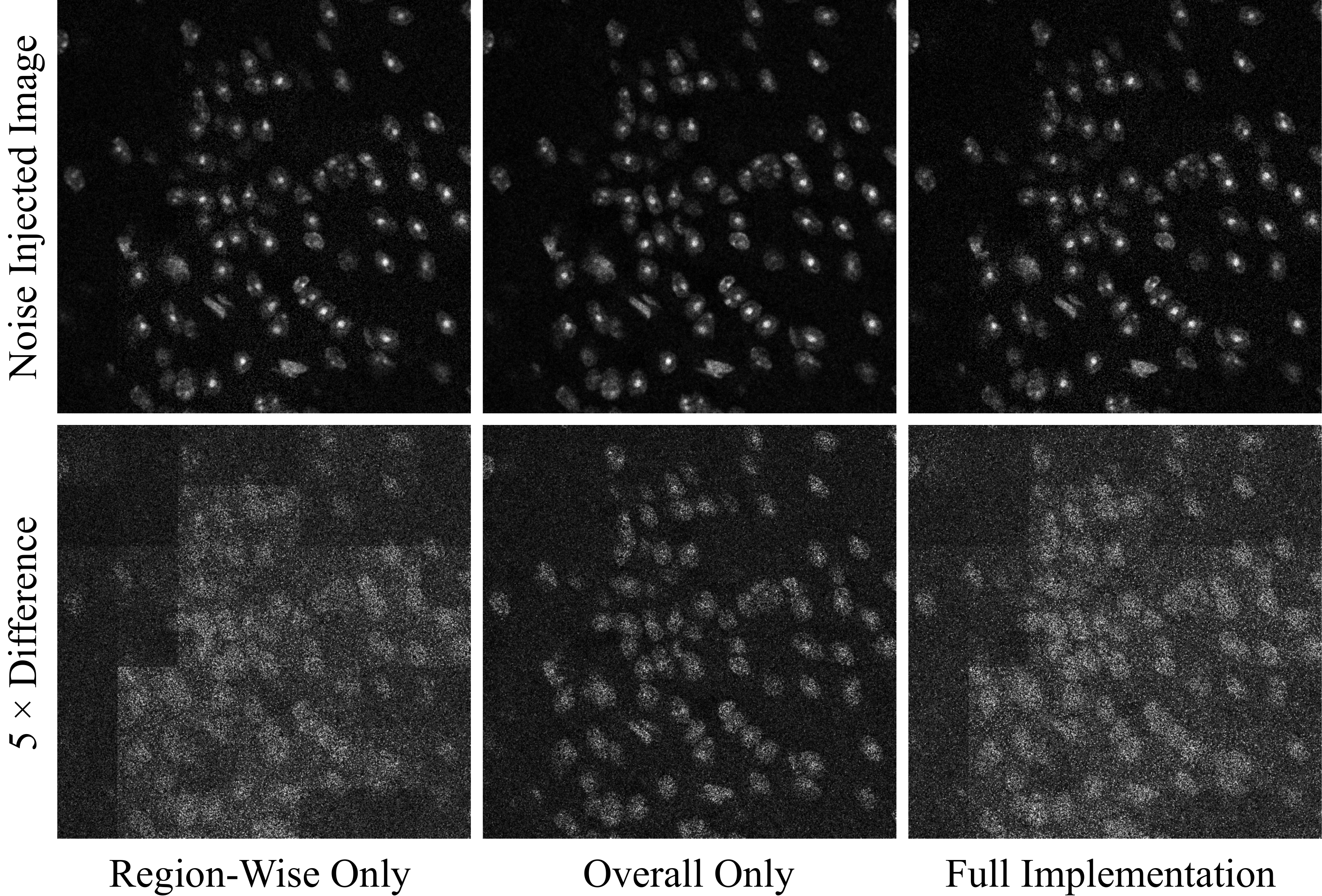}
    \caption{Visualization of Noise Injection components on FMI. The first raw is the image after Noise Injection, and the second row is $5\times$ difference between noisy image and ground truth, obtained with the same procedure of Fig. \ref{fig:NoiseObservation}.}
    \label{fig:NoiseInjection}
\end{figure}

The purpose of the Noise Injection module is to generate noisy samples that feature noise distribution similar to the real ones, serving as data augmentation executed multiple times. For this aim, we design \textit{Region-Wise Noise Injection} and \textit{Overall Noise Injection}. A visualization of these two schemes is depicted in Fig. \ref{fig:NoiseInjection}.

\subsubsection{Preprocess}
\noindent \textbf{Pre-Denoise.} In FM2S, Pre-Denoise serves two key roles: preparing input for Noise Injection and network training target. Directly applying Noise Injection to the original noisy images generates noisier images than real ones, which is against the consistency rule of sample augmentation. To address this problem, we introduce Pre-Denoise first to obtain an image with relatively less noise as the input of Noise Injection. In addition to Noise Injection, considering that the processed image is less corrupted with noise, it also serves as the training target for DNN, which is crucial to the whole pipeline. In consideration of efficiency and performance, we choose the simple median filter for Pre-Denoise.

\noindent \textbf{Channel Amplification.} To further enhance the network's generalization ability, we introduce Channel Amplification before the Noise Injection module, which increases the number of training samples with high parallel efficiency. This is done by first splitting the channels of the filtered input image and then repeating every channel for $\lambda $ times. Given a filtered image ${I_{filtered}} \in {\mathbb{R}^{H \times W \times C}}$, the amplified output is ${I_{amp}} \in {\mathbb{R}^{H \times W \times \lambda C}}$, where $\lambda $ denotes the amplification factor. 

\subsubsection{Region-Wise Noise Injection}
Following the observation in Section \ref{sec:Motivation&Observation}, we design Region-Wise Noise Injection, which applies adaptive MPG noise to regions of the image based on regional pixel intensity.

In this step, a mask indicating the region intensity of MPG is first generated. Specifically, the image is split into chunks, and the mask ${M_{[k]}}$ is the mean of pixel intensity in $k-th$ region. Subsequently, the Poisson distribution factor ${\lambda _{p[k]}}$ and Gaussian standard deviation ${\sigma _{g[k]}}$ for $k-th$ region are calculated by corresponding mask ${M_{[k]}}$ by
\begin{equation}
    \begin{aligned}
        {\sigma _{g[k]}} &:= {k_g} \times {M_{[k]}}, \\
        {\lambda _{p[k]}} &:= \frac{{{k_p}}}{{{M_{[k]}}}},
    \end{aligned}
\end{equation}
where ${k_g}$ and ${k_p}$ respectively denote the Gaussian mapping factor and the Poisson mapping factor, which are hyperparameters optimized for specific microscopes. After obtaining the noise distribution factor for each region, adaptive MPG will be added to each region.
\begin{equation}
    \begin{aligned}
    x_{i,j} &:= x_{i,j}^p + x_{i,j}^g, \\
    \lambda _{p[k]} x_{i,j}^p &\sim \mathcal{P}(\lambda _{p[k]} x_{i,j}), \\
    x_{i,j}^g &\sim \mathcal{N}(0, \sigma _{g[k]}^2),
    \end{aligned}
\end{equation}
where $x_{i,j}$ denotes $(i,j)$ pixel with the injection of adaptive noise in \textit{k}th region, $x_{i,j}^p$ denotes pixel with added Poisson noise and $x_{i,j}^g$ denotes Gaussian noise.

\subsubsection{Overall Noise Injection}
\cite{mannam2022real} points out that thermal noise caused by electronic components exists in the whole image, independent of the pixel intensity. Therefore, depending solely on Region-Wise is insufficient to simulate real noise to a certain extent. To address this problem, we add Poisson noise to the whole image, which introduces low-spatial-correlation noise to the image. The process can be modeled as

\begin{equation}
    \begin{aligned}
    {x_{i,j}} &:= x_{i,j}^p,\\
    {\lambda _p}x_{i,j}^p &\sim \mathcal{P}({\lambda _p}{x_{i,j}}),
    \end{aligned}
\end{equation}
where $x_{i,j}$ denotes $(i,j)$ pixel after Overall Noise Injection, $x_{i,j}^p$ denotes pixel with Poisson noise and ${\lambda _p}$ denotes the distribution factor for Poisson noise.

\subsection{Network and Training}
\label{sec:NN&Training}
\subsubsection{Network Architecture}
In single-image denoising, overfitting is more common due to the insufficient data, thus DNNs with millions of parameters may not achieve the optimal results. Inspired by \cite{mansour2023zero}, we confirm that a plain 3-layer DNN with $\sim$\textbf{3.5k} parameters is capable of striking a balance between efficiency and efficacy. The detailed network architecture is depicted in Fig. \ref{fig:Overall}, which comprises the convolution layer and Leaky ReLU activation.

\subsubsection{Training Strategy}
FM2S is trained in a coarse-to-fine manner, consisting of \textit{Coarse Feature Learning} and \textit{Fine-Grained Feature Learning}, both of which enable the denoising network to capture different levels of feature.

\noindent \textbf{Stage 1: Coarse Feature Learning.} This stage trains the DNN to learn the mapping from noisy image $y$ to the filtered image $u$. The training loss is

\begin{equation}
    {{\cal L}_{stage1}} = MSE({f_\theta }(y),u).
\end{equation}

This stage mainly serves as the initialization of the denoising network because the noisy image usually has less uncertainty than the synthetic image $u$, and thus, it takes much fewer steps than Stage 2 training.

\noindent \textbf{Stage 2: Fine-Grained Feature Learning.} This stage aims to enhance the generalization ability of the network and trains with abundant synthetic images, which enable the DNN to better learn the noise mapping. The loss is modelled as

\begin{equation}
    {{\cal L}_{stage2}} = MSE({f_\theta }(\hat \eta (u)),u),
\end{equation}

\noindent where $\hat \eta ( \cdot )$ denotes the Noise Injection module and $u$ for the filtered image. In view of the number of samples, this stage takes up the main proportion of training.

\section{Experiments}
In this section, we first present our experiment settings in Section \ref{sec:ExperimentSettings}. Comparisons between FM2S and baselines, along with study on different noise levels can be found in Section \ref{sec:MainResults}. Ablations are carried out in Section \ref{sec:AblationStudy}.

\begin{table*}[htbp]
\centering
\resizebox{\linewidth}{!}{
\begin{tabular}{cccccccccccc}
\toprule
\multirow{2}{*}{Category}& \multirow{2}{*}{Method} & \multirow{2}{*}{Source} & \multicolumn{2}{c}{Confocal} & \multicolumn{2}{c}{TwoPhoton} & \multicolumn{2}{c}{WideField} & \multicolumn{2}{c}{Weighted Avg.} & \multirow{2}{*}{Time} \\
\cmidrule(lr){4-5} \cmidrule(lr){6-7} \cmidrule(lr){8-9} \cmidrule(lr){10-11}
& & & PSNR & SSIM & PSNR & SSIM & PSNR & SSIM & PSNR & SSIM & \\
\midrule
\multirow{6}{*}{\rotatebox{90}{Impaired Dataset}}
& Restormer \cite{zamir2022restormer} & \textit{CVPR'22} & 33.81 & 0.911 & 31.75 & 0.872 & 27.45 & 0.504 & 31.53 & 0.796 & - \\
& MaskDenoising \cite{Chen_2023_CVPR} & \textit{CVPR'23} & 32.99 & 0.878 & 32.09 & 0.847 & 31.51 & 0.808 & 32.32 & 0.850 & - \\
& TransCLIP \cite{cheng2024transfer} & \textit{CVPR'24} & 32.55 & 0.730 & 30.66 & 0.838 & 27.10 & 0.485 & 30.56 & 0.705 & - \\
& AT-BSN \cite{chen2024exploring} & \textit{CVPR'24} & 33.09 & 0.866 & 31.82 & 0.841 & 32.02 & 0.790 & 32.40 & 0.838 & - \\
& AP-BSN$^{*}$ \cite{lee2022ap} & \textit{CVPR'22} & - & - & - & - & - & - & 31.99 & 0.836 & - \\
& CVF-SID$^{*}$ \cite{neshatavar2022cvf} & \textit{CVPR'22} & - & - & - & - & - & - & 32.73 & 0.843 & - \\
\midrule
\multirow{11}{*}{\rotatebox{90}{Single Image}} 
& BM3D \cite{dabov2007image} & \textit{TIP'07} & 31.28 & 0.820 & 32.20 & 0.879 & 31.26 & 0.760 & 31.58 & 0.824 & 7 sec \\ 
& DIP \cite{ulyanov2018deep} & \textit{CVPR'18} & 33.94 & 0.902 & 32.35 & 0.853 & 28.57 & 0.603 & 32.07 & 0.811 & 52 sec \\ 
& Noise2Self \cite{batson2019noise2self} & \textit{ICML'19} & \uline{34.75} & \textbf{0.922} & \uline{33.48} & \textbf{0.889} & 27.05 & 0.473 & 32.40 & 0.799 & 27 min \\ 
& ZSN2N \cite{mansour2023zero} & \textit{CVPR'23} & 34.27 & 0.908 & 32.76 & 0.869 & 25.84 & 0.398 & 31.66 & 0.768 & 13 sec \\ 
& MPI \cite{ma2024masked} & \textit{NIPS'24} & 34.35 & 0.906 & 32.60 & 0.863 & 31.08 & 0.726 & 32.95 & 0.846 & 25 sec \\ 
& MASH \cite{chihaoui2024masked} & \textit{CVPR'24} & 34.21 & 0.896 & 33.09 & 0.875 & \textbf{33.62} & \textbf{0.863} & \uline{33.69} & \uline{0.881} & 48 sec \\ 
& TTT-MIM \cite{mansour2024ttt} & \textit{ECCV'24} & 32.99 & 0.878 & 32.09 & 0.847 & 31.51 & 0.808 & 32.32 & 0.808 & \textbf{2 sec} \\ 
& AMSNet \cite{hu2024gaussianavatar} & \textit{ECCV'24} & 31.14 & 0.810 & 30.99 & 0.802 & \uline{33.17} & 0.833 & 31.60 & 0.813 & \textbf{2 sec} \\
& \cellcolor{gray!20} \textbf{FM2S} & \cellcolor{gray!20} \textbf{Ours} & \cellcolor{gray!20} \textbf{35.15} & \cellcolor{gray!20} \uline{0.911} & \cellcolor{gray!20} \textbf{33.67} & \cellcolor{gray!20} \uline{0.882} & \cellcolor{gray!20} 33.10 & \cellcolor{gray!20} \uline{0.850} & \cellcolor{gray!20} \textbf{34.15} & \cellcolor{gray!20} \textbf{0.886} & \cellcolor{gray!20} \uline{6 sec} \\ 
& Self2Self$^{*}$ \cite{quan2020self2self} & \textit{CVPR'20} & - & - & - & - & - & - & 30.76 & 0.695 & - \\
& ScoreDVI$^{*}$ \cite{cheng2023score} & \textit{ICCV'23} & - & - & - & - & - & - & 33.10 & 0.865 & - \\
\bottomrule
\end{tabular}
}
\caption{Denoising performance of FM2S and baselines on the test set of FMD dataset \cite{zhang2019poisson} at raw noise level. `Impaired Dataset' denotes those trained with SIDD dataset \cite{abdelhamed2018high} and tested on the FMD test set, and `Single Image' denotes zero-shot methods that are trained and evaluated with the same image. The time results for single-image denoising methods include training and inference. The weighted average is calculated by averaging the results according to image proportion. The results with * are reported from \cite{chihaoui2024masked}.}
\label{table:MainResults}
\end{table*}

\begin{table*}[htbp]
\centering
\footnotesize
\begin{tabular}{ccccccccccc}
\toprule
Metric & Noisy & TTT-MIM \cite{mansour2024ttt} & BM3D \cite{dabov2007image} & MPI \cite{ma2024masked} & ZSN2N \cite{mansour2023zero} & AMSNet\cite{hu2024gaussianavatar} & N2S \cite{batson2019noise2self} & DIP \cite{ulyanov2018deep} & MASH \cite{chihaoui2024masked} & \cellcolor{gray!20}FM2S (Ours) \\ 
\midrule
PSNR   & 26.75 & 25.99 & 28.02 & 28.17 & 28.20 & 28.30 & 28.36 & 28.44 & \uline{28.89} & \cellcolor{gray!20}\textbf{28.93} \\ 
SSIM   & 0.636 & 0.855 & 0.860 & 0.885 & 0.883 & 0.878 & \textbf{0.919} & 0.898 & \uline{0.905} & \cellcolor{gray!20}\textbf{0.919} \\ 
\bottomrule
\end{tabular}
\caption{Denoising performance comparison for FM2S and other single-image methods on SRDTrans \cite{li2023spatial} dataset. The best results of approaches trained with a single image are in \textbf{bold}, and the second best are \uline{underlined}.}
\label{table:SRDTrans_results}
\end{table*}

\begin{figure*}[htbp]
    \centering
    \begin{subfigure}{\textwidth}
        \centering
        \includegraphics[width=\textwidth]{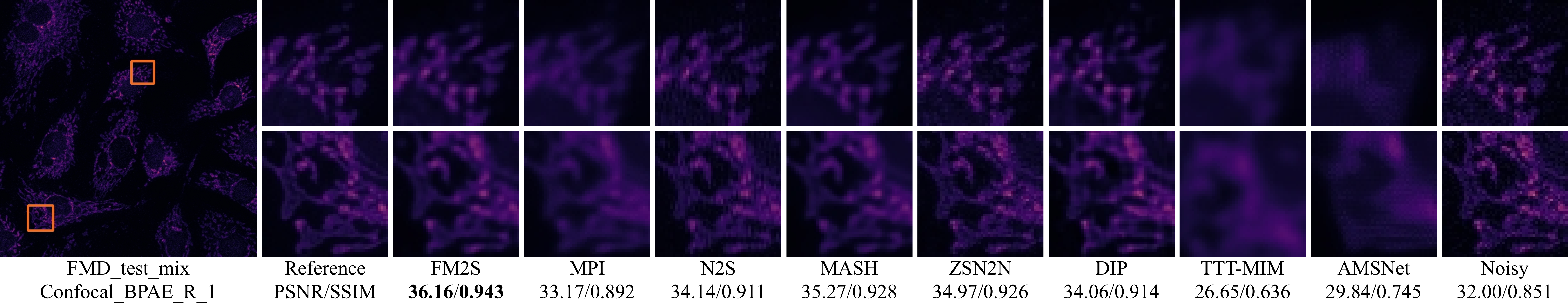}
        \label{fig:Confocal}
    \end{subfigure}
    \hfill
    \begin{subfigure}{\textwidth}
        \centering
        \includegraphics[width=\textwidth]{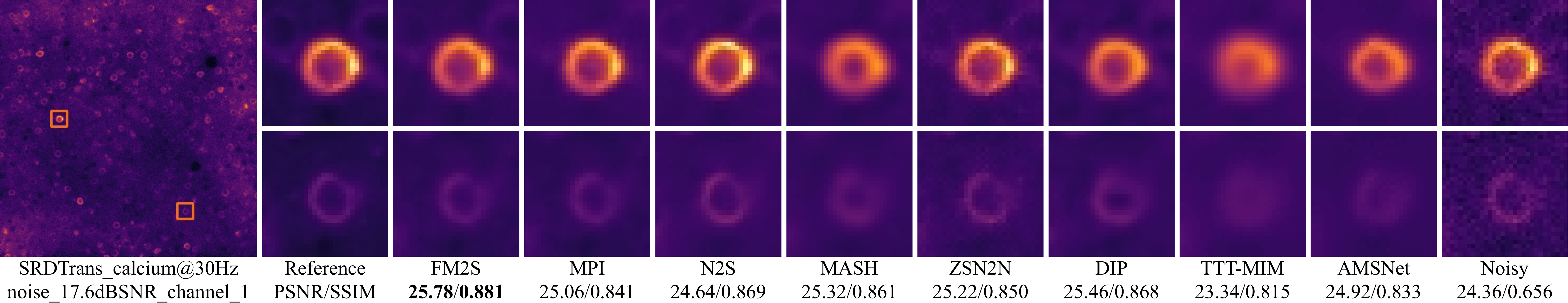}
        \label{fig:SRDTrans}
    \end{subfigure}
    \caption{Qualitative visualization of our method and other comparisons. The upper is a fluorescence micrograph of mouse brain tissue captured by a Confocal microscope from the FMD dataset \cite{zhang2019poisson}, and the lower is one of simulated calcium imaging data sampled at 30 Hz from the SRDTrans dataset \cite{li2023spatial}. The raw images are in grayscale, and we transform them into inferno color for better visualization.}
    \label{fig:Effects}
\end{figure*}

\begin{table*}[htbp]
\centering
\resizebox{\linewidth}{!}{
\begin{tabular}{lcccccccccccc}
\toprule
\multirow{2}{*}{Level} & \multirow{2}{*}{Noisy} & \multicolumn{8}{c}{Single Image} & \multicolumn{3}{c}{Impaired Dataset} \\
\cmidrule(lr){3-10}\cmidrule(lr){11-13}
& & BM3D & DIP & MPI & N2F & ZSN2N & TTT-MIM & MASH & FM2S & TransCLIP & AT-BSN & Restormer \\ 
\midrule
avg1   & 27.217 & 31.581 & 32.071 & 32.949 & 32.628 & 31.662 & 32.3179 & \uline{33.692} & \textbf{34.146} & 30.556 & 32.398 & 31.535 \\ 
avg2  & 30.078 & 33.740 & 33.374 & 34.155 & 32.977 & 33.249 & 33.0391 & \uline{34.143} & \textbf{35.058} & 33.231 & 33.337 & 33.444 \\
avg4  & 32.861 & \uline{36.007} & 34.854 & 35.585 & 35.851 & 35.332 & 33.4017 & 35.134 & \textbf{36.825} & 35.694 & 34.335 & 35.375 \\
avg8  & 36.027 & \uline{37.744} & 36.050 & 36.581 & 37.740 & 37.705 & 33.8414 & 35.756 & \textbf{38.366} & 37.370 & 35.314 & 37.330 \\
avg16 & 39.703 & 39.370 & 36.897 & 37.714 & 40.099 & \textbf{40.852} & 34.2714 & 36.517 & \uline{40.745} & 39.170 & 36.133 & 39.354 \\
\bottomrule
\end{tabular}
}
\caption{The PSNR results on various noise levels. Higher avg\textit{N} values indicate less noise corruption. The best results are in \textbf{bold}, and the second best are in \uline{underlined}.}
\label{table:NoiseLevel}
\end{table*}

\begin{table*}
\footnotesize
    \centering
    \begin{minipage}[t]{0.73\textwidth}
    \centering
    \begin{tabular}{l|ccccccccc}
    \toprule
             & TTT-MIM & AMSNet & MASH & MPI & DIP & N2F & N2S & ZSN2N & \cellcolor{gray!20}FM2S \\ \midrule
    Params   & 7.80M & 2.23M & 0.99M & 744.7k & 572.7k & 259.2k & 223.1k & 22.3k & \cellcolor{gray!20}\textbf{3.5k} \\
    FLOPs    & 101.5G & 7.0T & 90.0G & 63.8G & 3.5G & 135.9G & 116.9G & 11.2G & \cellcolor{gray!20}\textbf{1.8G} \\ \bottomrule
    \end{tabular}
    \caption{Computation consumption for FM2S and baselines. We display trainable network parameter size (Parameter Size) and floating-point operations (FLOPs) with an input size $512\times512\times1$. Each method is under default settings and network architecture.}
    \label{table:Computational Efficiency}
    \end{minipage}
    \hfill
    \begin{minipage}[t]{0.26\textwidth}
    \centering
    \begin{tabular}{l|cc}
    \toprule
             & w/ NI & w/o NI \\
    \midrule
    w/ TT  & \textbf{34.15} & 33.62 \\
    w/o TT & 33.04 & 31.46 \\ 
    \bottomrule
    \end{tabular}
    \caption{PSNR results for ablation of Median Filter for Noise Injection and Training Target.}
    \label{tab:AblateFilter}
    \end{minipage}
\end{table*}

\subsection{Experimental Settings}
\label{sec:ExperimentSettings}
\noindent \textbf{Dataset.}
We adopt two public datasets for FMI denoising: the Fluorescence Microscopy Denoising (FMD) dataset \cite{zhang2019poisson} and the SRDTrans dataset \cite{li2023spatial}. The FMD dataset includes three commonly used fluorescence microscopes: Confocal, TwoPhoton and WideField and provides images at five different noise levels, achieved by averaging $N$ noisy images with $N \in \{ 1,2,4,8,16\}$. The SRDTrans dataset contains calcium imaging and single-molecule localization microscopy data; each image has hundreds of channels. For evaluation, we focus on calcium imaging data sampled at 30 Hz with a TwoPhoton microscope and choose first 10 channels to build a subset.

Single-image-based methods are trained and tested on one image, while methods that rely on vast data to train are directly evaluated with FMI using their released weights for real-noise removal, which are trained with real-noise datasets like SIDD \cite{abdelhamed2018high} or synthetic-noise datasets. Therefore, we did not split the extra training set and took the FMD test set along with the SRDTrans subset for evaluation.

\noindent \textbf{Baseline.}
We compare FM2S with various denoising methods, including those that rely on single images for both training and inference, as well as methods trained on large datasets. For Single-image-based methods, we choose BM3D \cite{dabov2007image} as a representation of traditional methods. Regarding those relying on DNNs, we choose AMSNet \cite{hu2024gaussianavatar}, MASH \cite{chihaoui2024masked}, MPI \cite{ma2024masked}, TTT-MIM \cite{mansour2024ttt}, ZSN2N \cite{mansour2023zero}, Noise2Self\cite{batson2019noise2self}, Deep Image Prior (DIP) \cite{ulyanov2018deep}, Self2Self \cite{quan2020self2self} and ScoreDVI \cite{cheng2023score}. All the baselines are under the default experiment settings for real noise removal, which is the corresponding option for FMI noise. For TTT-MIM, the original adaption steps setting is 8. However, this is insufficient for FMI denoising, so we increase it to 100.


Regarding methods that rely on large datasets, we choose three recent representative works with high generalization ability: AT-BSN \cite{Chen_2024_CVPR}, TransCLIP \cite{cheng2024transfer} and MaskDenoising \cite{Chen_2023_CVPR}. Additionally, we report the results of Restormer \cite{zamir2022restormer}, AP-BSN \cite{lee2022ap} and CVF-SID \cite{neshatavar2022cvf}. Each method is evaluated with officially released weights and codes for real noise denoising.

\noindent \textbf{Implementation Details.}
In our implementation, we utilize the Adam optimizer \cite{kingma2014adam} to update the network, with an initial learning rate of 0.001. The amplification factor $\lambda$ is 6 for the SRDTrans dataset and 2 for the FMD dataset. The basic configuration of $k_{gau}$ and $k_{poi}$ for Region-Wise Noise Injection is 200 and 30, respectively. As for Overall Noise Injection, the Poisson noise distribution factor ${\lambda _p} = 60$. Detailed configurations and learning hyperparameters can be found in the supplementary material. All the experiments are completed on one NVIDIA RTX 3090 GPU.

\subsection{Main Results}
\label{sec:MainResults}
\subsubsection{Evaluation on Real Fluorescence Microscopies}

We evaluated FM2S on real noisy FMI from the FMD \cite{zhang2019poisson} and SRDTrans \cite{li2023spatial} datasets, focusing on the raw noise level. Since most supervised and self-supervised methods take much longer to train than single-image-based methods, we do not report their time consumption. Different amplification factors $\lambda$ lead to slightly different times, so we report the average time consumption for the whole test set. The denoising performance and time consumption results are in Table \ref{table:MainResults} and Table \ref{table:SRDTrans_results}.

For Confocal images, FM2S achieves the highest PSNR and second highest SSIM. For TwoPhoton, despite slightly falling short of Noise2Self \cite{batson2019noise2self}, FM2S finishes denoising \textbf{270}$\boldsymbol{\times}$\textbf{faster}. While for WideField images, which are not well registered \cite{lin2023self}, FM2S ranks second among all the single-image-based methods. For the SRDTrans dataset, FM2S achieves outstanding effects both in PSNR and SSIM, indicating its strong generalization ability to multiple datasets. Besides, w.r.t. average performance, FM2S is comparable to the current SOTA method MASH \cite{chihaoui2024masked}, but $8\times$ faster. Although FM2S is not as fast as TTT-MIM \cite{mansour2024ttt}, it delivers significantly better denoising results within an acceptable time limit. This proves that FM2S strikes a balance between denoising performance and efficiency.

Notably, most methods trained on a large dataset, including those for enhanced generalization ability (e.g. MaskDenoising \cite{Chen_2023_CVPR}, TransCLIP \cite{cheng2024transfer}), fail to remove the intricate noise in FMI, suggesting the severe out-of-distribution problem while denoising FMI. Fine-tuning these models with FMI image pairs may address this problem. However, as mentioned in Section \ref{sec:intro}, obtaining ground truth is difficult and time-consuming. This highlights the urgency of single-image-based methods for FMI denoising.

Figure \ref{fig:Effects} visually compares FM2S with other methods. FM2S preserves more fine details like edges and highlights of the cell, producing clearer denoising results. More visualizations can be found in supplementary material.

\subsubsection{Evaluation on Different Noise Level}


We examine our method's performance with FMI at different noise levels. We additionally examine Noise2Fast \cite{lequyer2022fast} with its hyperparameters following the original settings. The results are shown in Table \ref{table:NoiseLevel}. FM2S consistently exhibits outstanding performance across all noise levels and especially does well in images with relatively low noise levels. It is notable that the raw images in the FMD dataset are acquired with excitation laser power as low as possible to generate noisy samples \cite{zhang2019poisson}, which means that in most conditions, bio-researchers deal with \textit{images with lower noise levels}, highlighting the practical value of FM2S.


\subsubsection{Computational Efficiency}



In addition to denoising performance, we further evaluate the network parameter size and computational consumption of FM2S. All experiments are conducted on the same platform described in Section \ref{sec:ExperimentSettings} with amplification factor $\lambda=1$. For methods with multiple network choices, we take the default configuration. Since traditional denoising methods do not benefit from a GPU and usually take much less time to conduct, we do not report their results. Results on time consumption are shown in Table \ref{table:MainResults} and analyses on DNNs are shown in Table \ref{table:Computational Efficiency}. Benefitting from a small network, FM2S features the fewest network parameters and FLOPs. Additionally, the small network requires much fewer optimization steps, enabling fast denoising within 6 seconds. 

\subsection{Ablation Studies}
\label{sec:AblationStudy}

\begin{table}[t]
\centering
\setlength{\tabcolsep}{8pt}
\resizebox{\linewidth}{!}{
\begin{tabular}{lcccc}
\toprule
\multirow{2}{*}{Ablation Option} & \multicolumn{2}{c}{FMD} & \multicolumn{2}{c}{SRDTrans} \\ 
\cmidrule(lr){2-3} \cmidrule(lr){4-5}
 & PSNR & SSIM & PSNR & SSIM \\ 
\midrule
(a) w/o Region           & 32.41          & 0.795          & 28.62          & 0.895          \\
(b) w/o Overall          & 33.89          & 0.876          & 28.65         & 0.910          \\
(c) w/o Channel+         & 34.02          & 0.882          & 28.68         & 0.900          \\
(d) w/ Stage 1 Only      & 23.75          & 0.681          & 25.59          & 0.780          \\
(e) w/ Stage 2 Only      & 34.08          & 0.884          & 28.82          & 0.916          \\ \midrule \rowcolor{gray!20}
(f) Full Conduction              & \textbf{34.15} & \textbf{0.886} & \textbf{28.93} & \textbf{0.919} \\
\bottomrule
\end{tabular}
}
\caption{Ablation study of Noise Injection components and training stages in FM2S. The `w/o' stands for `without,' and `/w' stands for `with.'}
\label{table:AblationStudy}
\end{table}

\subsubsection{Noise Injection Components}
The Noise Injection module is composed of Region-Wise Noise Injection and Overall Noise Injection, which takes channel-amplified images and generates samples for network training. We ablated the main components of Noise Injection and present the results in Table \ref{table:AblationStudy}. In experiments detailed in lines (a) and (b), removing Region-Wise Noise Injection will significantly damage the performance, which agrees with our observation. In line (c), we find that channel amplification is essential for Confocal and TwoPhoton images. Still, it has limited effects on WideField images, so it is not applied in the original implementation for WideField.

\subsubsection{Pre-Denoise}
\noindent \textbf{Design Choices.} Pre-Denoise is done by applying an elementary denoising method to the noisy image. Here, we compare filters, NLM \cite{buades2005non}, BM3D \cite{dabov2007image} for Pre-Denoise and report their effects with time consumption. The results are shown in Table \ref{tab:DesignCoices}, a conclusion can be drawn that the simple Median Filter can achieve both performance and speed.

\noindent \textbf{Effects.} We removed Pre-Denoise for Noise Injection (NI) and Training Target (TT), and the results in Table \ref{tab:AblateFilter} prove that removing either of them would lead to suboptimal results. We hypothesize that this is caused by the broken consistency of Eq. \ref{eq:optimize with filtered} when the filter is ablated because the Noise Injection must receive images $u$ that share some common features with ground truth $x$, like concentration in the spectrum, which is done by accomplished by Pre-Denoise.

\subsubsection{Influences of The Hyperparameters}

\begin{figure}[tbp]
    \centering
    \begin{subfigure}[b]{0.235\textwidth}
        \centering
        \includegraphics[width=\textwidth]{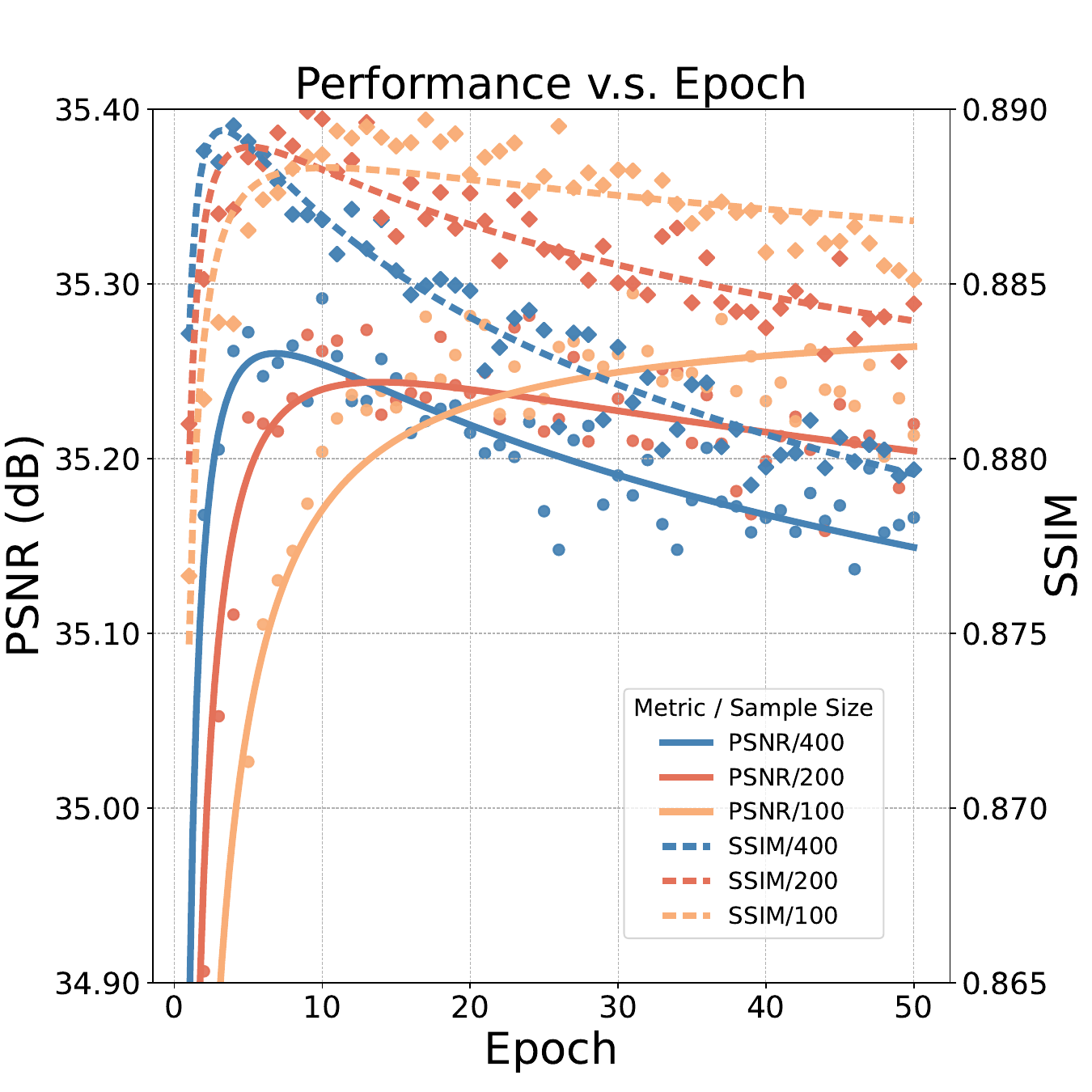}
    \end{subfigure}
    \hfill
    \begin{subfigure}[b]{0.235\textwidth}
        \centering
        \includegraphics[width=\textwidth]{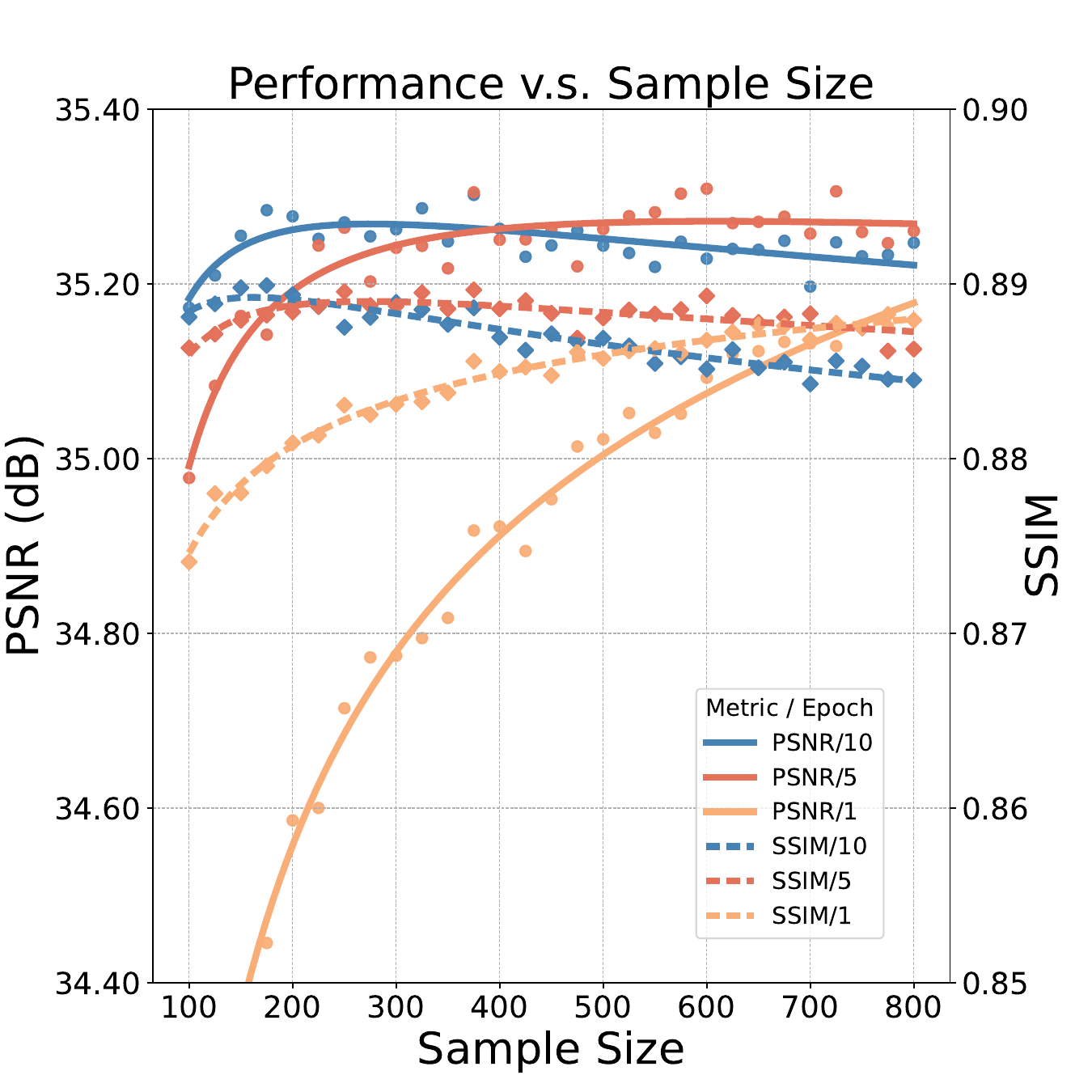}
    \end{subfigure}
    \caption{Denoising performance with various Epoch and Sample Size on a randomly chosen subset. Each dot is the average of 5 independent results to reduce random errors, with detailed raw data in the supplementary material.}
    \label{fig:KeyHypermeters}
\end{figure}

Intuitively, the denoising performance relies on the size of training samples (Sample Size) and training epochs (Epoch). As Fig. \ref{fig:KeyHypermeters} shows, the performance initially improves rapidly with increasing epochs but eventually declines due to potential overfitting. In contrast, the performance improves steadily with a larger sample size. We choose 5 and 450 as the default Epoch and Sample Size to strike a balance between effects and speed.

\begin{table}[t]
\centering
\setlength{\tabcolsep}{8pt}
\resizebox{\linewidth}{!}{
\begin{tabular}{lcccc} \toprule
Metric & \cellcolor{gray!20} Median & Gaussian & NLM     & BM3D    \\ \midrule
PSNR(dB)   & \cellcolor{gray!20} \textbf{34.15} & 32.21  & 32.85 & 33.23 \\
SSIM & \cellcolor{gray!20} \textbf{0.886} & 0.845 & 0.869 & 0.789 \\
Time(s)   & \cellcolor{gray!20} 6.0 &  6.1  & \textbf{5.9} & 13.1 \\ \bottomrule
\end{tabular}
}
\caption{Design choices of Pre-Denoise. `Median' and `Gaussian' denote Median Filter and Gaussian Filter with window size 3.}
\label{tab:DesignCoices}
\end{table}

\subsubsection{Two-Stage Training}
FM2S features a two-stage training, where each stage serves a different function. We conduct ablation studies on the two stages and show the corresponding results in lines (d) and (e) in Table \ref{table:AblationStudy}. It is notable that Stage 2 contributes more to the overall performance than Stage 1. The reason is plausibly that Stage 2 takes much more steps than Stage 1, and Stage 2 learns more fine-grained details, which are more crucial for denoising.
\section{Conclusion}
In this paper, we propose FM2S, a novel zero-shot denoising framework tailored for fluorescence microscopy images that require no additional data beyond a noisy image. Guided by the physical noise statics, we fully exploit the inherent attributes of fluorescence microscopy noise and specifically design a Noise Injection module for robust data augmentation, thereby mitigating the out-of-distribution problem. An ultra-lightweight network is trained in 2 stages to gradually capture features from coarse noise priors to high-frequency details while achieving a balance between efficacy and efficiency. Our method has been verified to produce denoising results comparable with other SOTA denoisers under a wide variety of noise levels while consuming much less time. FM2S is suitable for practical applications, and we believe it will contribute to advancements in biology and medicine studies.

{
    \small
    \bibliographystyle{ieeenat_fullname}
    \bibliography{main}
}


\newpage
\maketitlesupplementary

\section{Extensive Visualizations}
\begin{figure*}[h]
\centering
    \begin{subfigure}[b]{\textwidth}
        \centering
        \includegraphics[width=\textwidth]{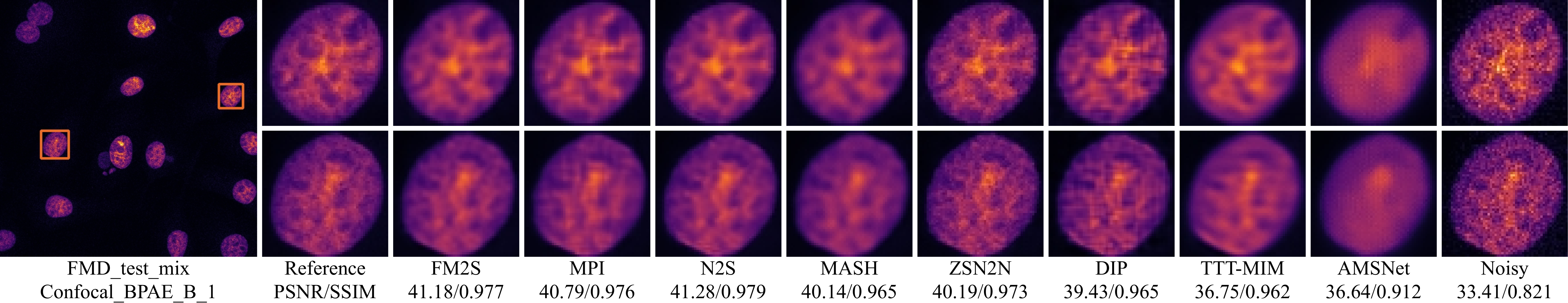}
    \end{subfigure}
        \hfill
    \begin{subfigure}[b]{\textwidth} 
        \centering
        \includegraphics[width=\textwidth]{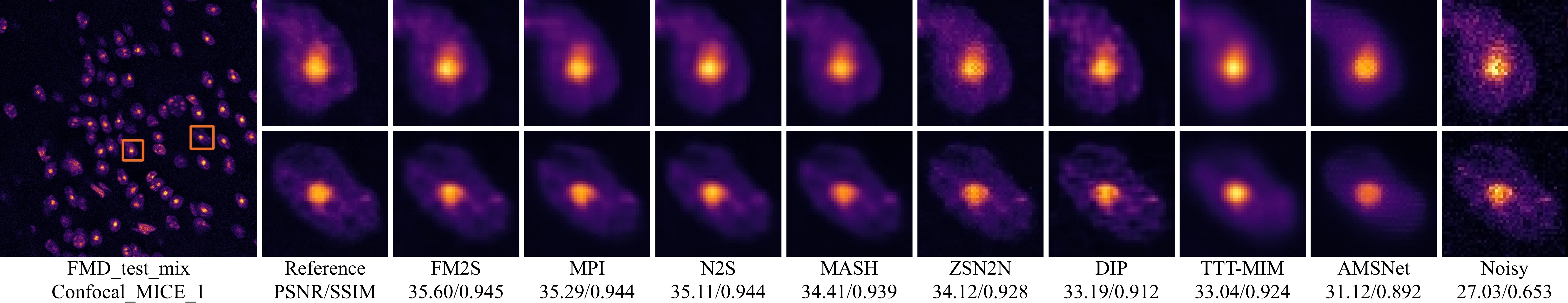}
    \end{subfigure}
        \hfill
    \begin{subfigure}[b]{\textwidth} 
        \centering
        \includegraphics[width=\textwidth]{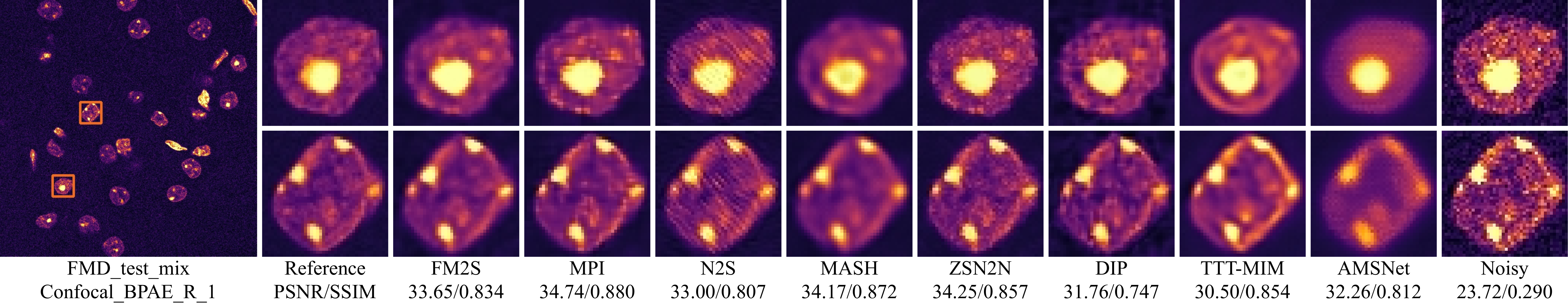}
    \end{subfigure}
    \caption{More visualizations for the denoising performance of FM2S and other methods. The raw images are in grayscale, and we transform them into inferno color for better visualization.}
    \label{fig:more_vis}
\end{figure*}

\begin{figure*}[htbp]
    \centering
    \begin{subfigure}[b]{0.49\textwidth}
        \centering
        \includegraphics[width=\textwidth]{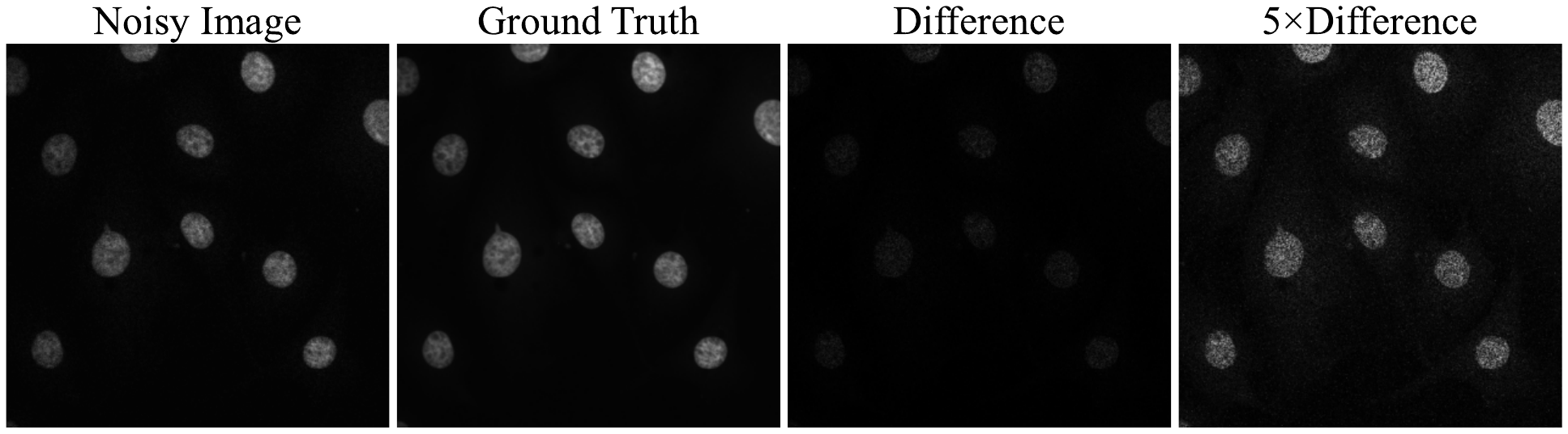}
    \end{subfigure}
    \hfill
    \begin{subfigure}[b]{0.49\textwidth}
        \centering
        \includegraphics[width=\textwidth]{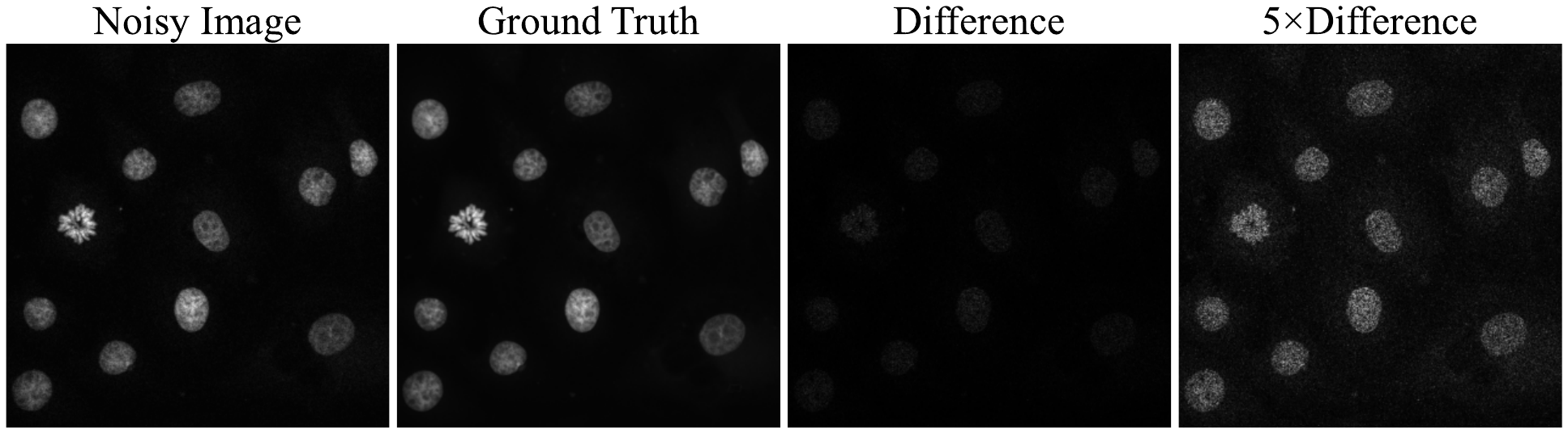}
    \end{subfigure}
    \hfill
    \begin{subfigure}[b]{0.49\textwidth}
        \centering
        \includegraphics[width=\textwidth]{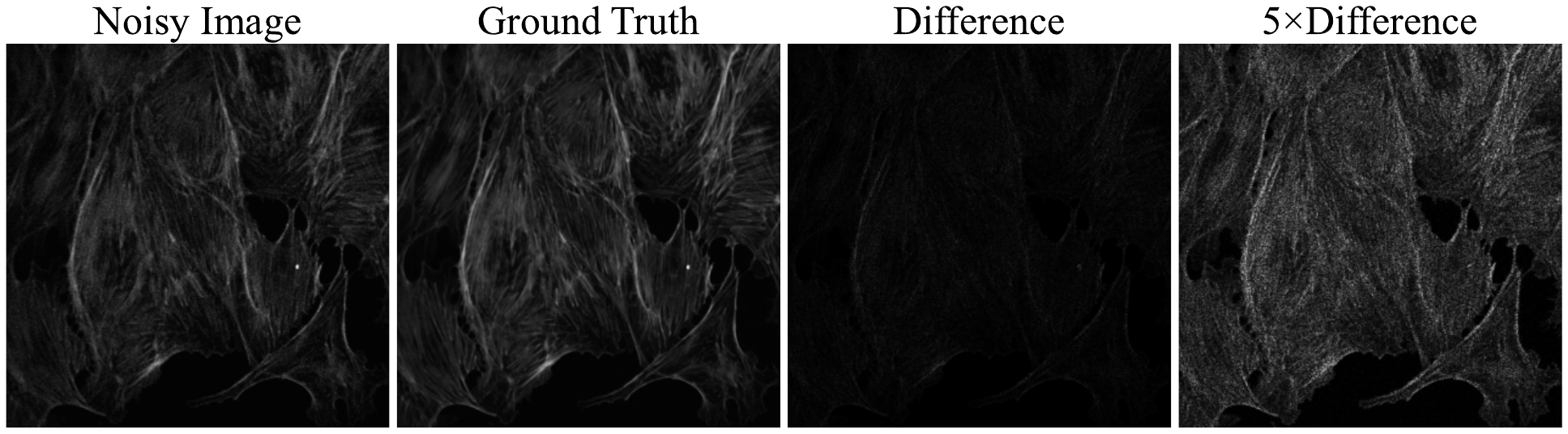}
    \end{subfigure}
    \hfill
    \begin{subfigure}[b]{0.49\textwidth}
        \centering
        \includegraphics[width=\textwidth]{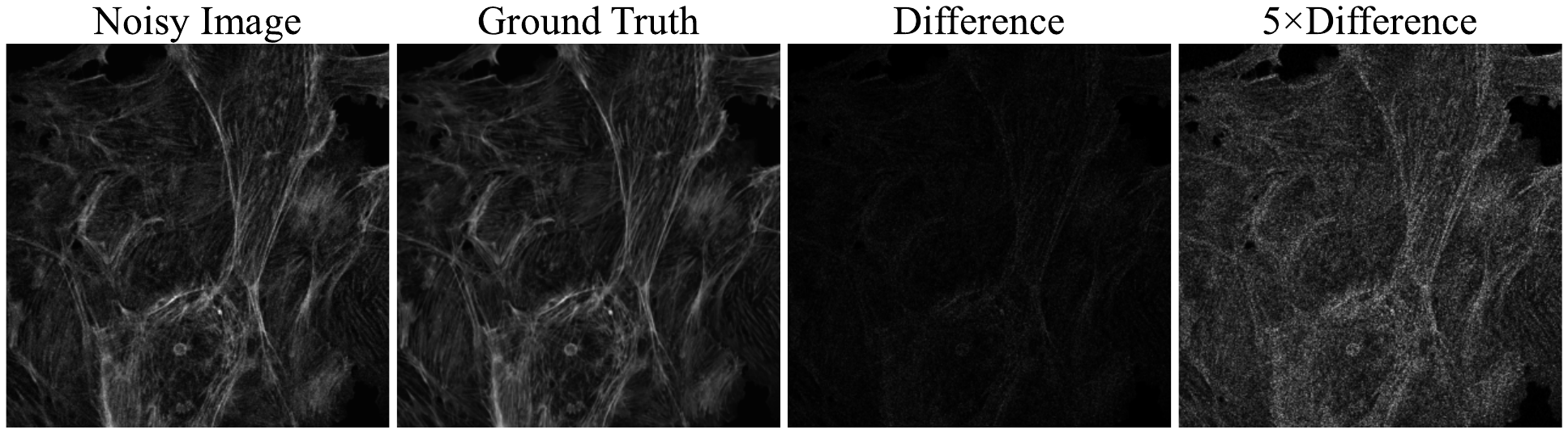}
    \end{subfigure}
    \hfill
    \begin{subfigure}[b]{0.49\textwidth}
        \centering
        \includegraphics[width=\textwidth]{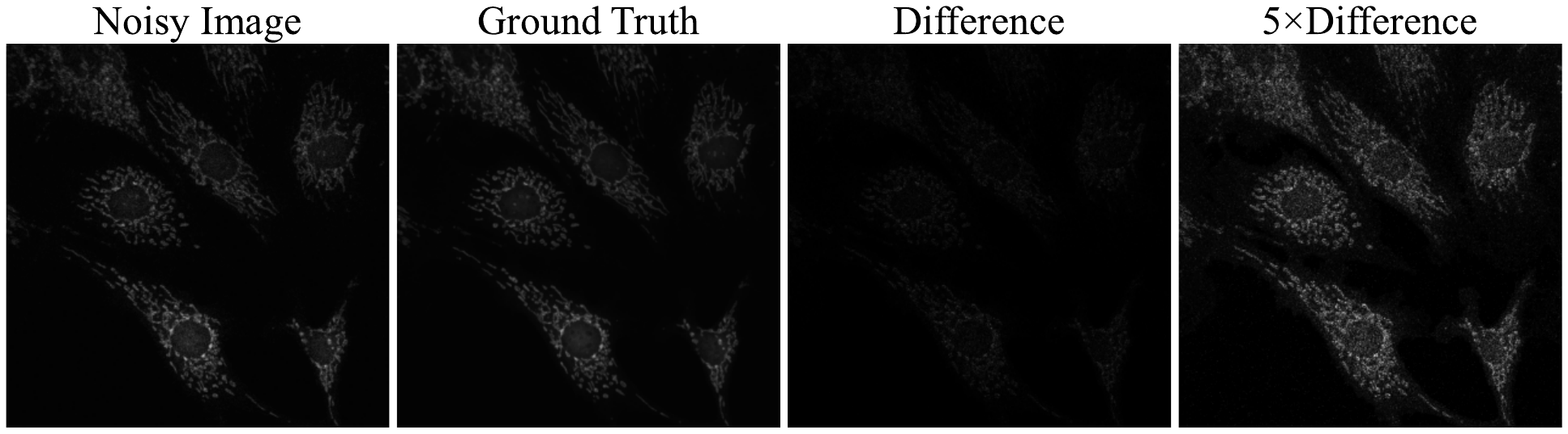}
    \end{subfigure}
    \hfill
    \begin{subfigure}[b]{0.49\textwidth}
        \centering
        \includegraphics[width=\textwidth]{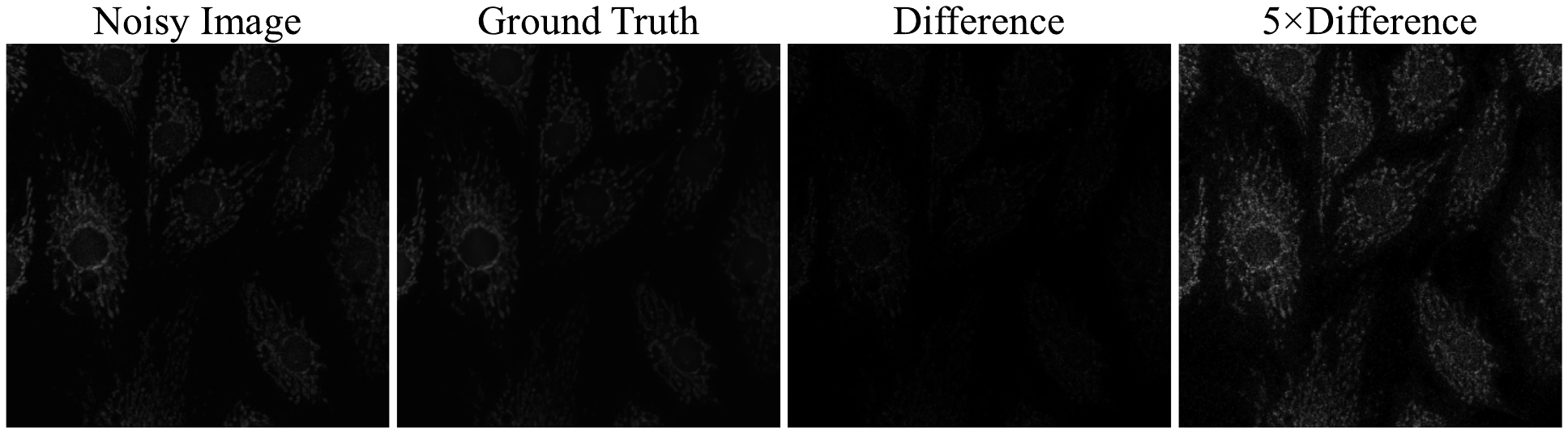}
    \end{subfigure}
    \hfill
    \begin{subfigure}[b]{0.49\textwidth}
        \centering
        \includegraphics[width=\textwidth]{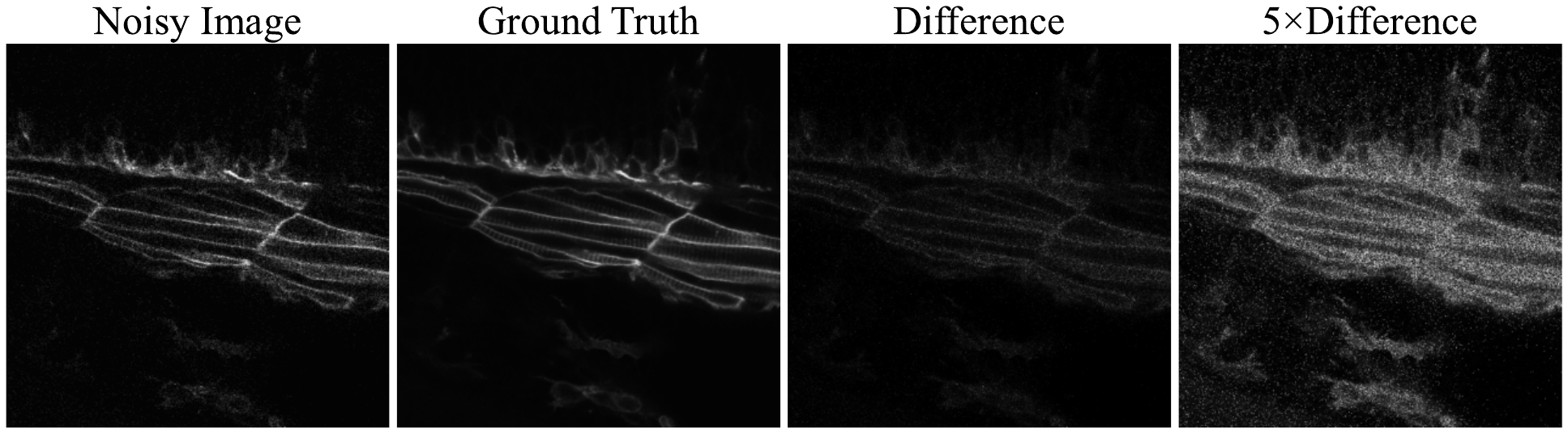}
    \end{subfigure}
    \hfill
    \begin{subfigure}[b]{0.49\textwidth}
        \centering
        \includegraphics[width=\textwidth]{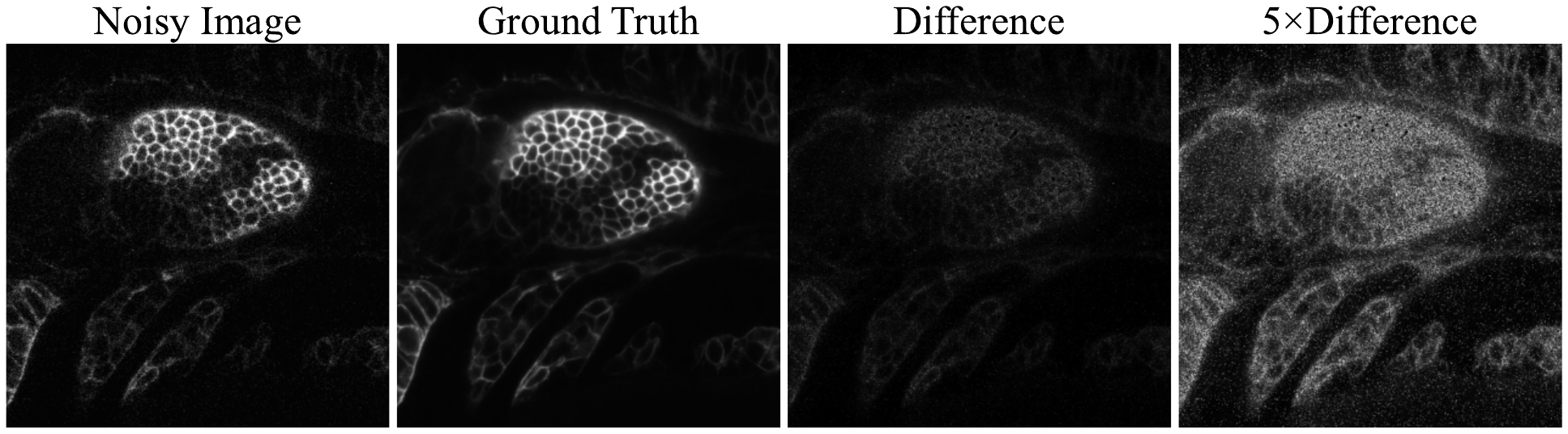}
    \end{subfigure}
    \hfill
    \begin{subfigure}[b]{0.49\textwidth}
        \centering
        \includegraphics[width=\textwidth]{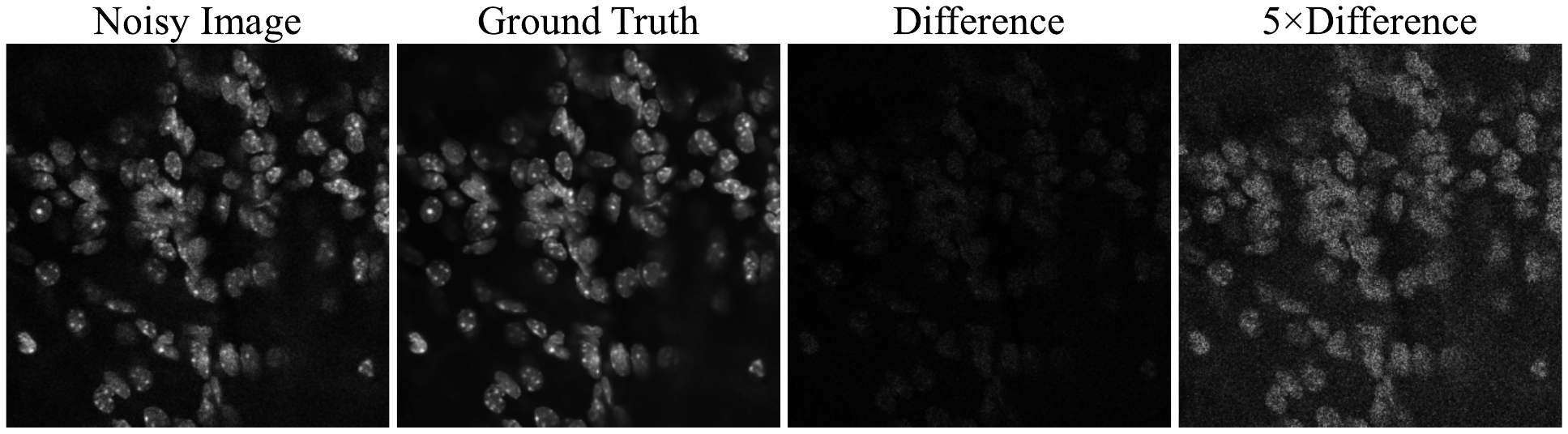}
    \end{subfigure}
    \hfill
    \begin{subfigure}[b]{0.49\textwidth}
        \centering
        \includegraphics[width=\textwidth]{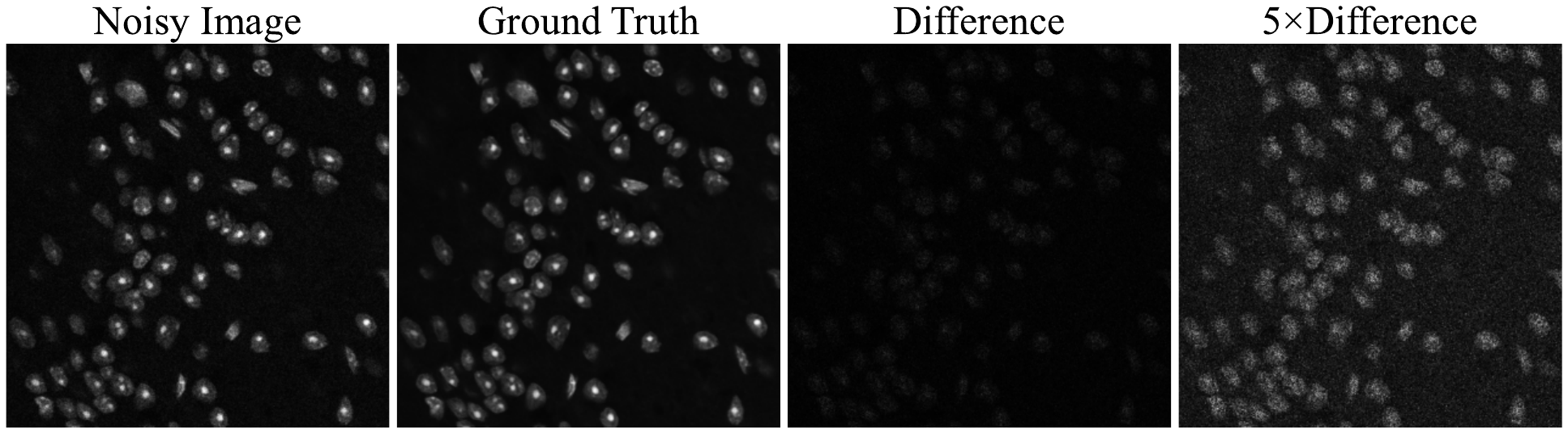}
    \end{subfigure}
    \hfill
    \begin{subfigure}[b]{0.49\textwidth}
        \centering
        \includegraphics[width=\textwidth]{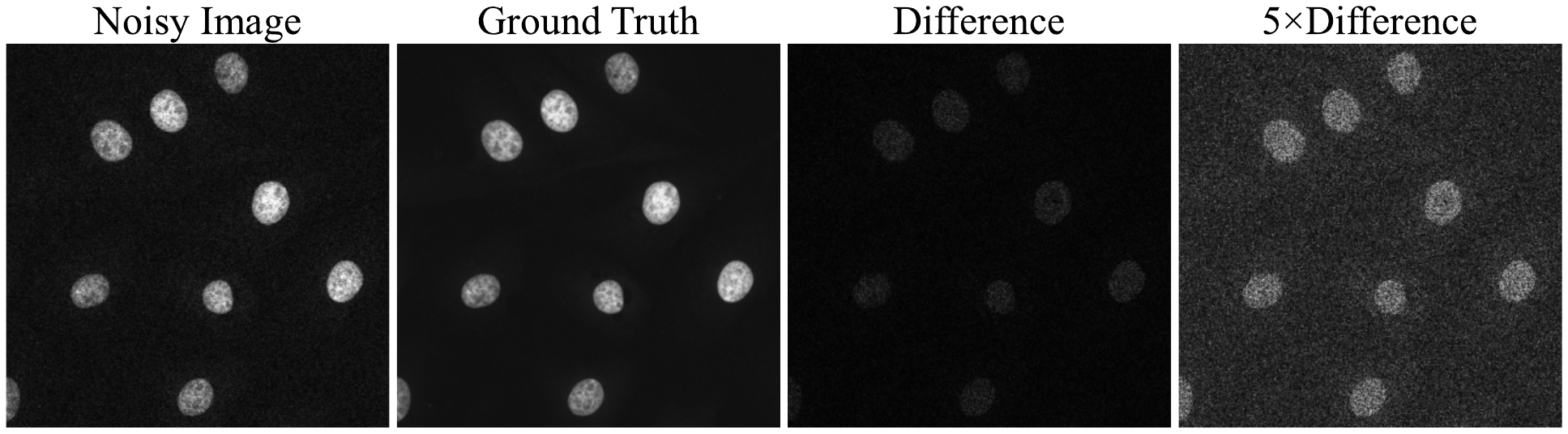}
    \end{subfigure}
    \hfill
    \begin{subfigure}[b]{0.49\textwidth}
        \centering
        \includegraphics[width=\textwidth]{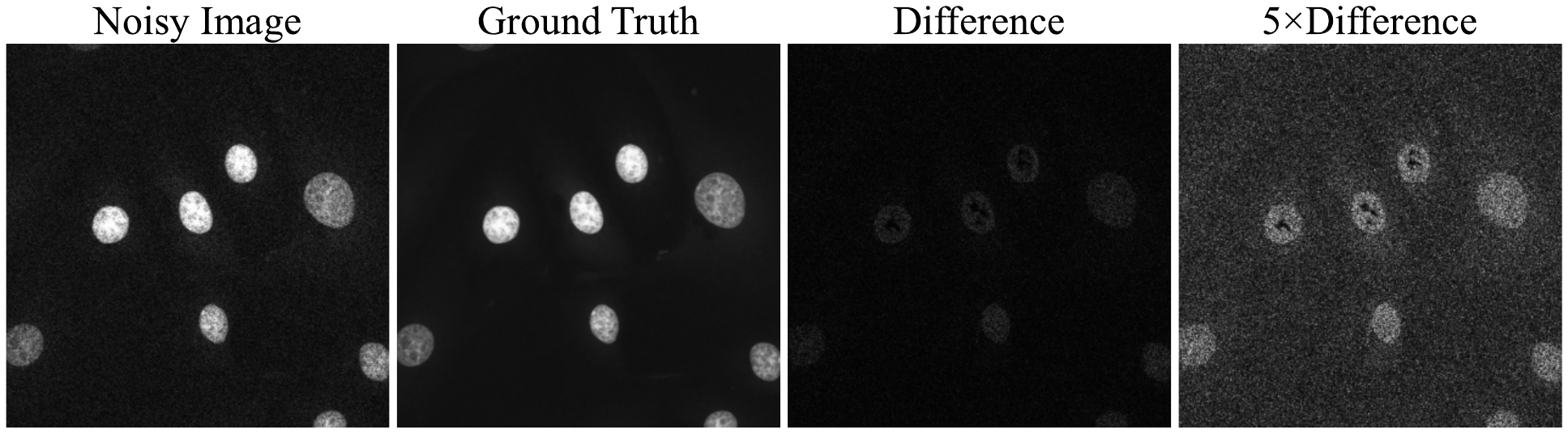}
    \end{subfigure}
    \hfill
    \begin{subfigure}[b]{0.49\textwidth}
        \centering
        \includegraphics[width=\textwidth]{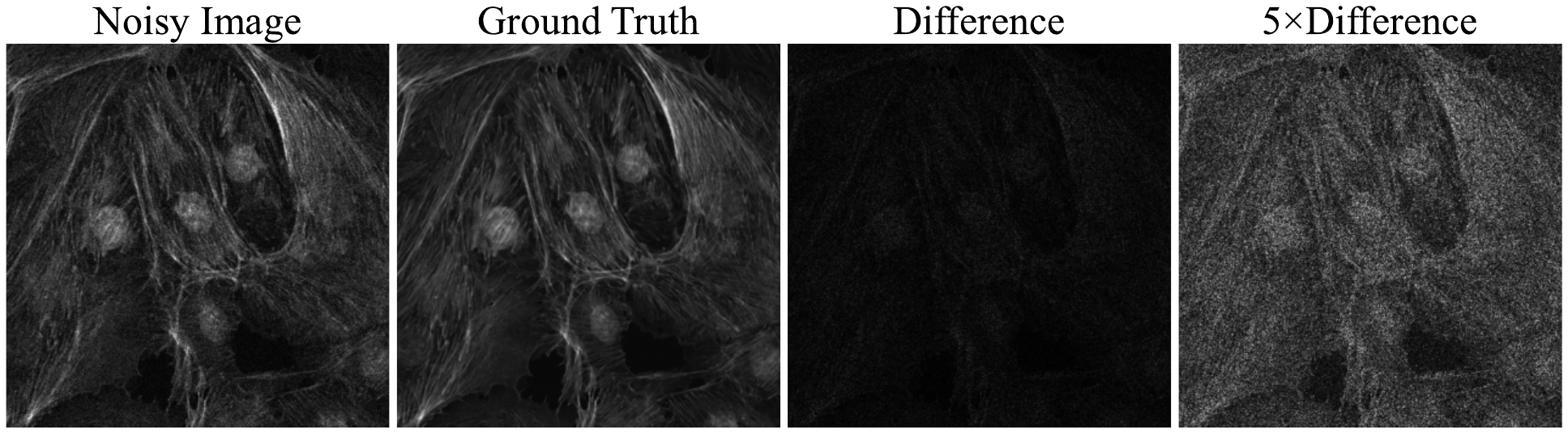}
    \end{subfigure}
    \hfill
    \begin{subfigure}[b]{0.49\textwidth}
        \centering
        \includegraphics[width=\textwidth]{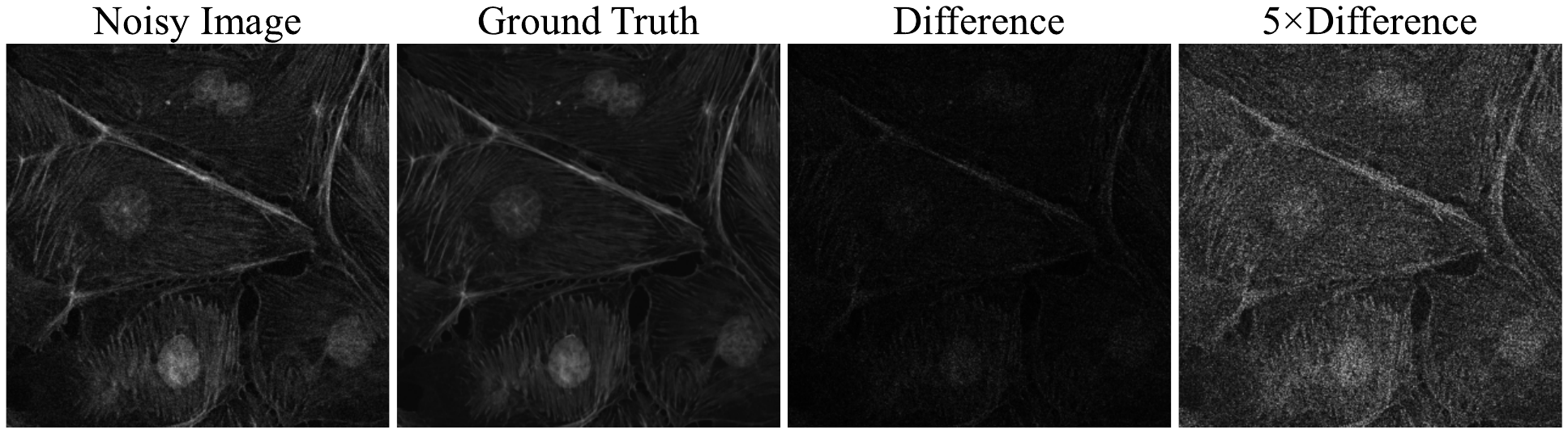}
    \end{subfigure}
    \hfill
    \begin{subfigure}[b]{0.49\textwidth}
        \centering
        \includegraphics[width=\textwidth]{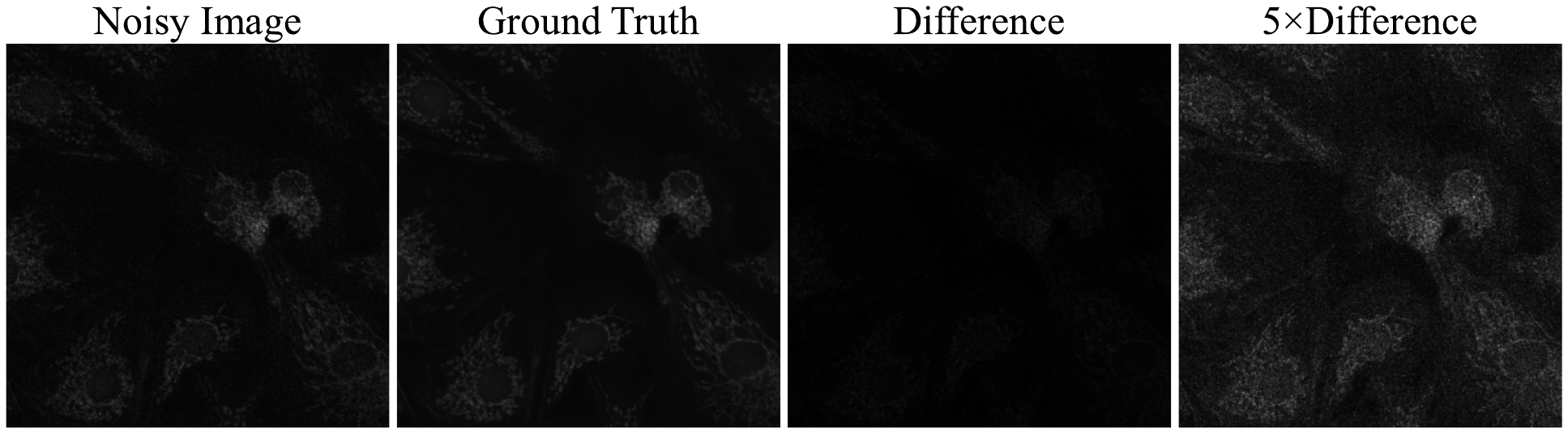}
    \end{subfigure}
    \hfill
    \begin{subfigure}[b]{0.49\textwidth}
        \centering
        \includegraphics[width=\textwidth]{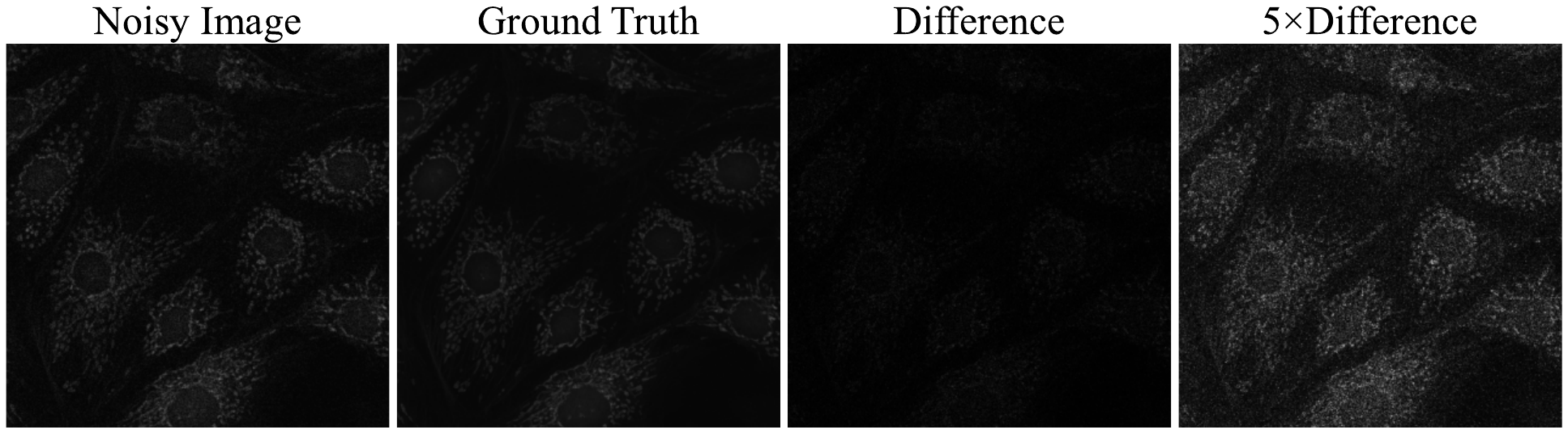}
    \end{subfigure}
    \hfill
    \begin{subfigure}[b]{0.49\textwidth}
        \centering
        \includegraphics[width=\textwidth]{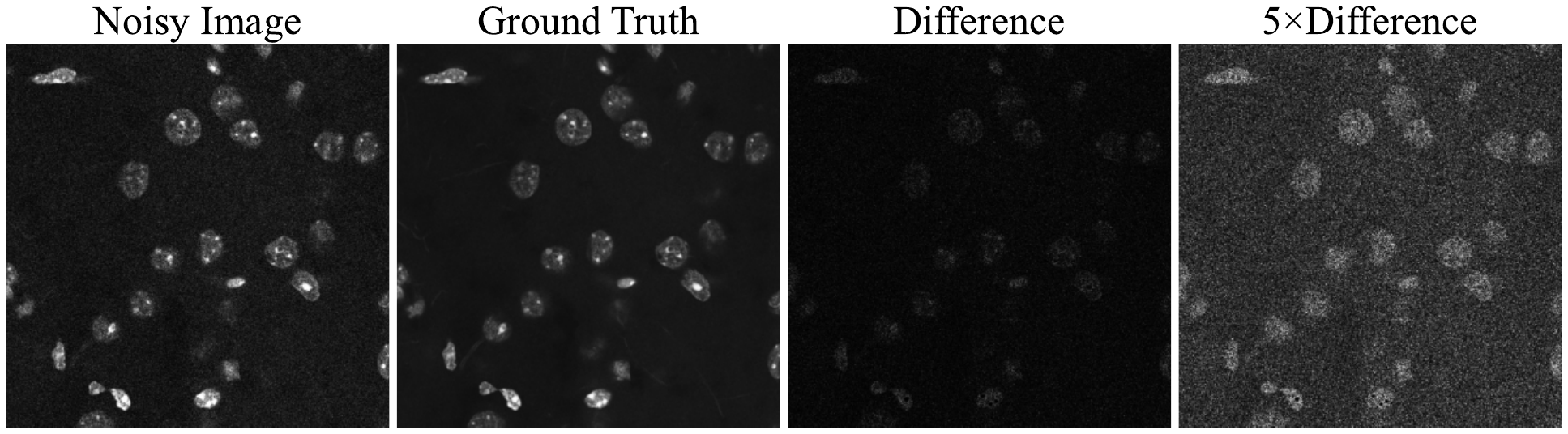}
    \end{subfigure}
    \hfill
    \begin{subfigure}[b]{0.49\textwidth}
        \centering
        \includegraphics[width=\textwidth]{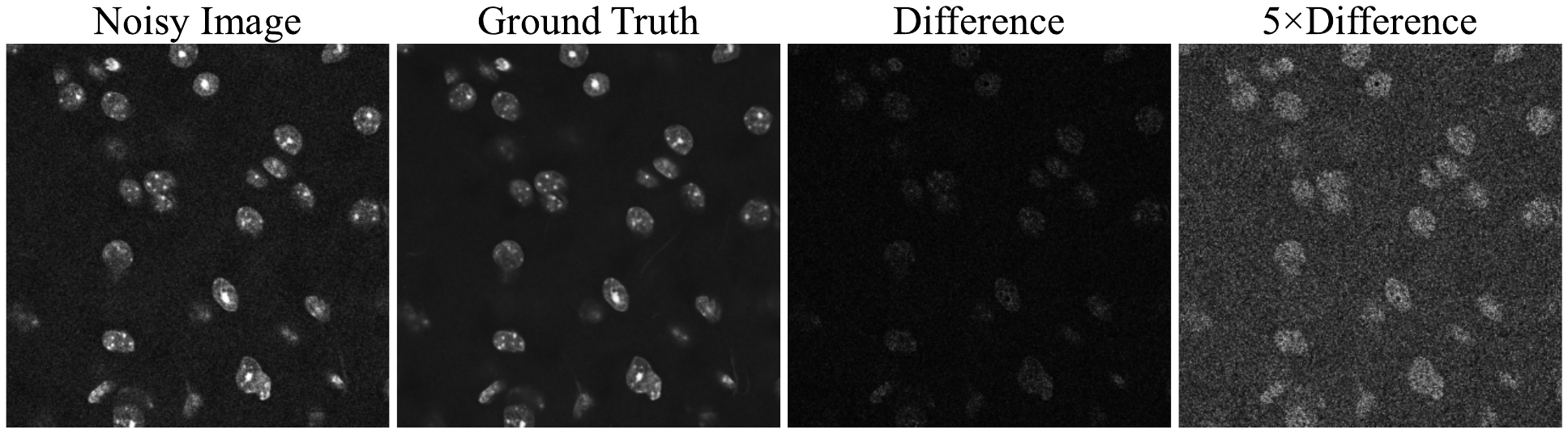}
    \end{subfigure}
    \caption{More visualizations for the difference between noisy images and corresponding ground truth.}
    \label{fig:MoreNoiseVisualization}
\end{figure*}

\begin{itemize}
    \item \textbf{Denoising effects.} Fig. \ref{fig:more_vis} shows more visualizations on various images. Although FM2S does not outperform all baselines in some examples, it reverses more crucial image details and agrees with human perception.
    \item \textbf{Noise Observation.} Fig. \ref{fig:MoreNoiseVisualization} depicts more examples of the difference between noisy FMI and ground truth, along with their 5$\times$ brightened version. Most images share a similar pattern described in Section \ref{sec:Motivation&Observation}, which is the foundation of Noise Injection.
\end{itemize}

\section{Limitations}
While showing great potential in FMI denoising, some limitations still exist in our method.
\begin{itemize}
    \item \textbf{Lack of Benchmark Dataset.} Although fluorescence microscopy is a widely used technology, datasets specifically designed for fluorescence microscopy image denoising are scarce. As mentioned in Section \ref{sec:intro}, constructing abundant noisy/clean image pairs for FMI is mostly impractical, limiting the evaluation of denoising performance.
    \item \textbf{Hyperparameters Optimization.} As Fig. \ref{fig:KeyHypermeters} depicts, denoising performance might go down slightly with growing Epoch and Sample Size, posing a need for an appropriate iteration number, which is similar to DIP \cite{ulyanov2018deep}. Besides, some hyperparameters need to be manually fine-tuned for optimal performance.
    \item \textbf{No Benefit from More Data.} Almost every single-image-based method suffers from a common issue. In cases where sufficient noisy/image pairs are available, methods that rely on one image do not benefit from the abundant data and usually fall short of common methods requiring vast data to train. Although this case is usually unachievable for FMI denoising, a stronger denoiser can still be built at the expense of vast resources and time.
\end{itemize}

\section{Discussions}
\subsection{Amplification Factor}
\begin{figure*}[htbp]
    \centering
    \begin{subfigure}[b]{0.235\textwidth}
        \centering
        \includegraphics[width=\textwidth]{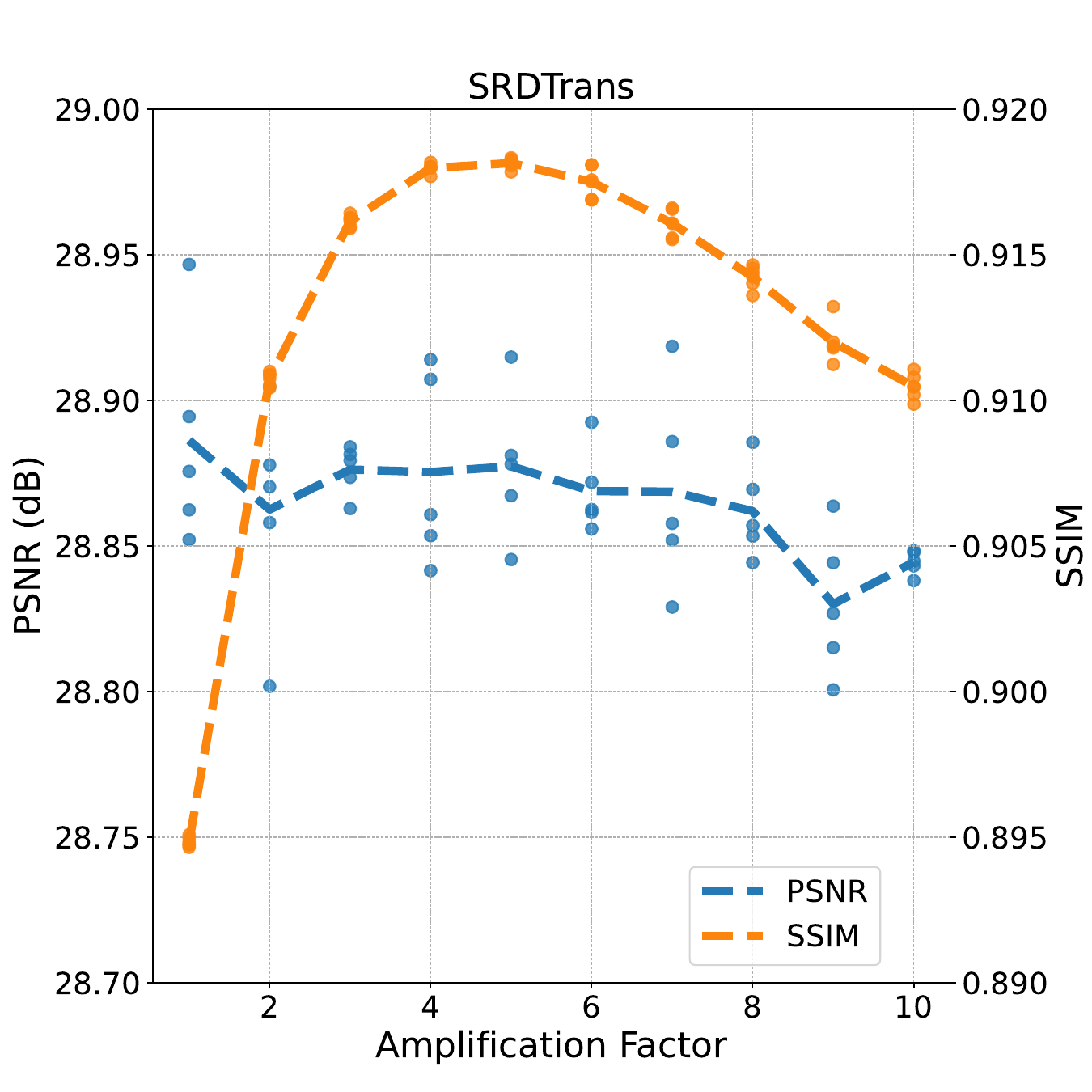}
    \end{subfigure}
    \hfill
    \begin{subfigure}[b]{0.235\textwidth}
        \centering
        \includegraphics[width=\textwidth]{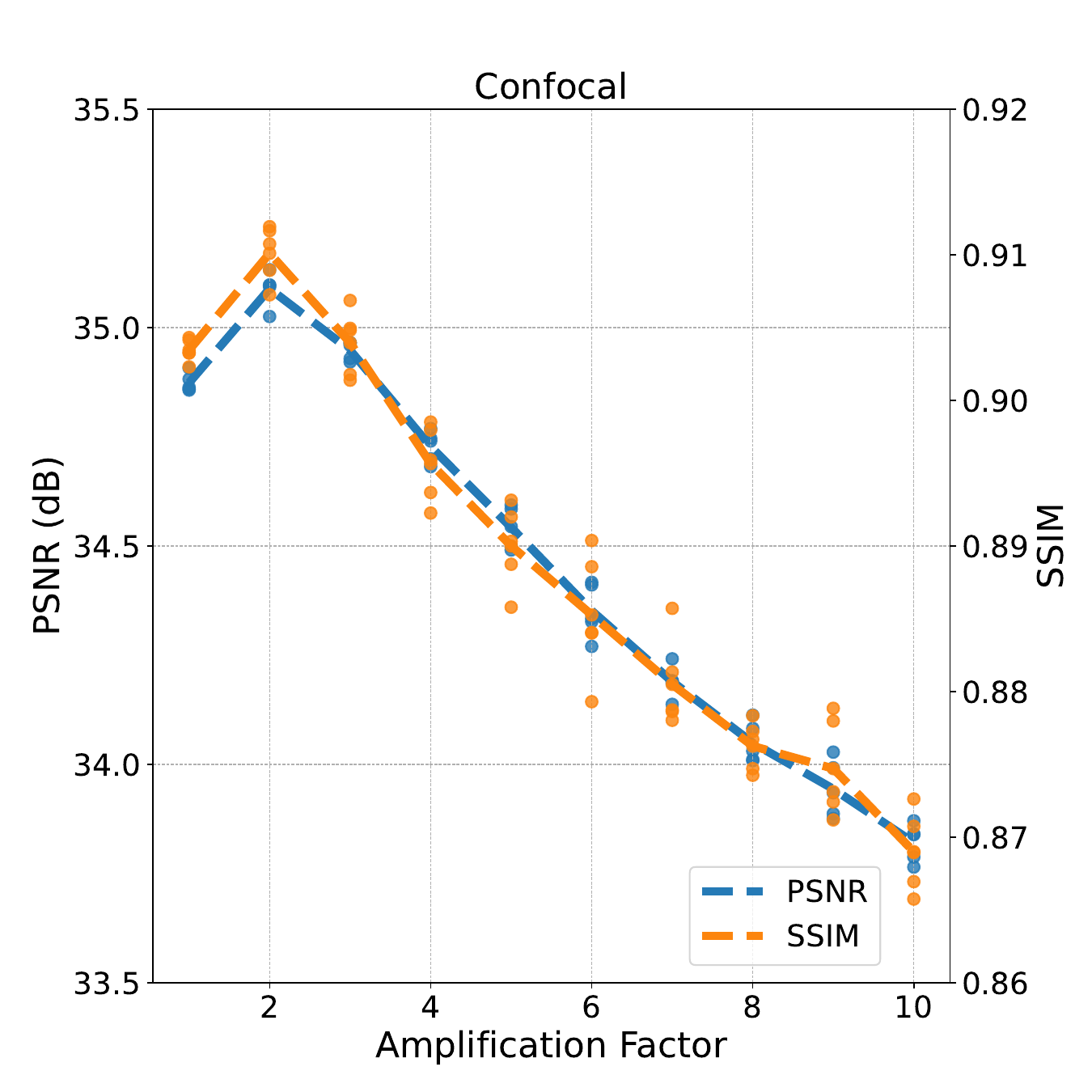}
    \end{subfigure}
        \hfill
    \begin{subfigure}[b]{0.235\textwidth}
        \centering
        \includegraphics[width=\textwidth]{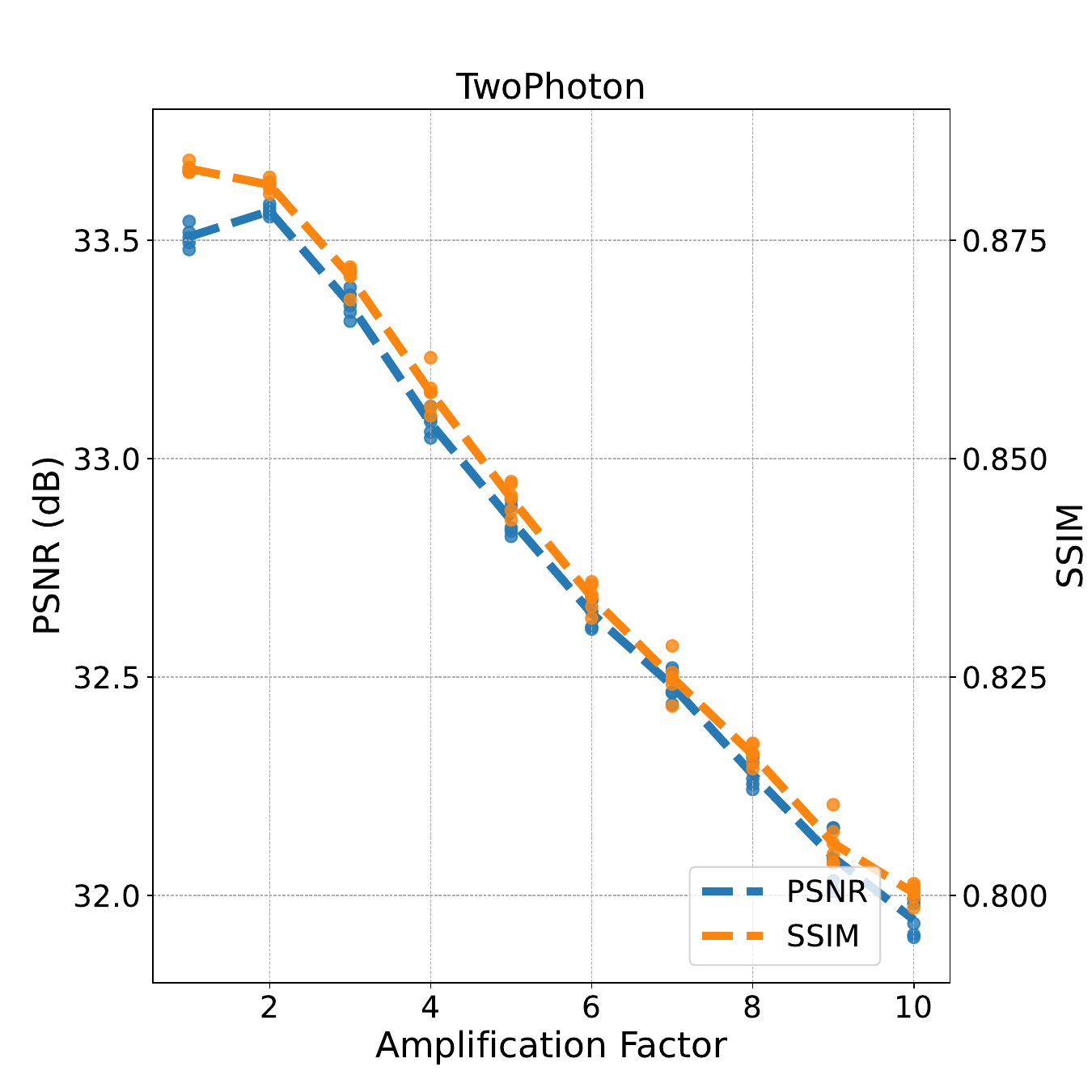}
    \end{subfigure}
        \hfill
    \begin{subfigure}[b]{0.235\textwidth}
        \centering
        \includegraphics[width=\textwidth]{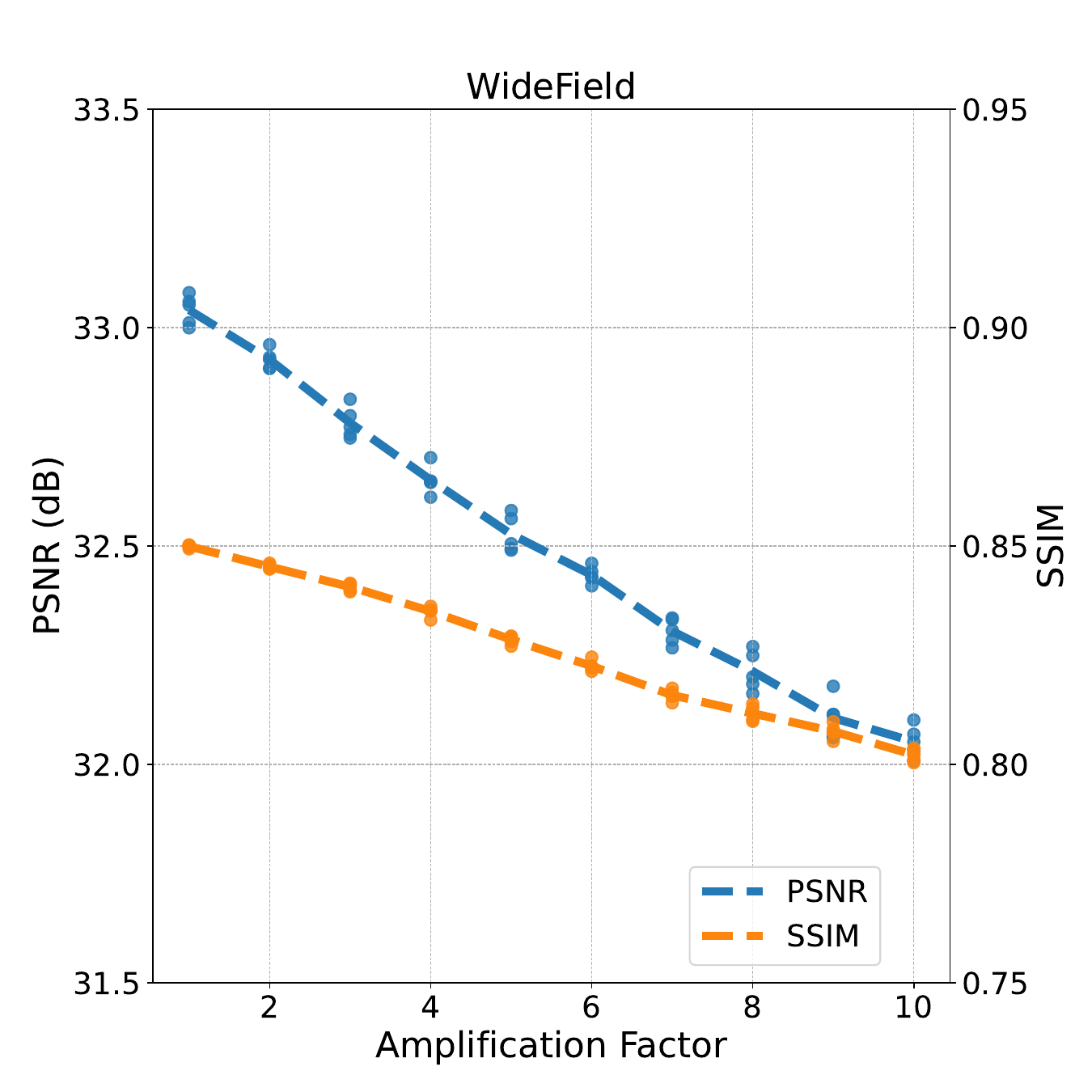}
    \end{subfigure}
    \caption{Denoising performance with various amplification factors. Each set is conducted 5 times independently to reduce random errors, and the line is plotted with the average of each set.}
    \label{fig:amp_lambda}
\end{figure*}

The distinct noise characteristics of different microscope types necessitate tailored amplification factors $\lambda$ for different microscope types to achieve optimal results. To investigate this relationship, we set a range of amplification factors for different microscopes and datasets under the same experiment settings and present the results in Fig \ref{fig:amp_lambda}. For the FMD dataset, denoising performance improves initially with increasing $\lambda$ but declines beyond a threshold in Confocal and TwoPhoton microscopes. At the same time, it continuously drops in WideField microscopes, indicating no beneficial range for higher amplification. For the SRDTrans dataset, the performance increases with $\lambda$ at first and goes down after an optimal value, which shows a similar trend to Confocal and TwoPhoton microscopes.

\subsection{Sample Size and Epoch for Training}
\begin{figure*}[h]
    \centering
    \begin{subfigure}{0.31\textwidth}
        \includegraphics[width=\linewidth]{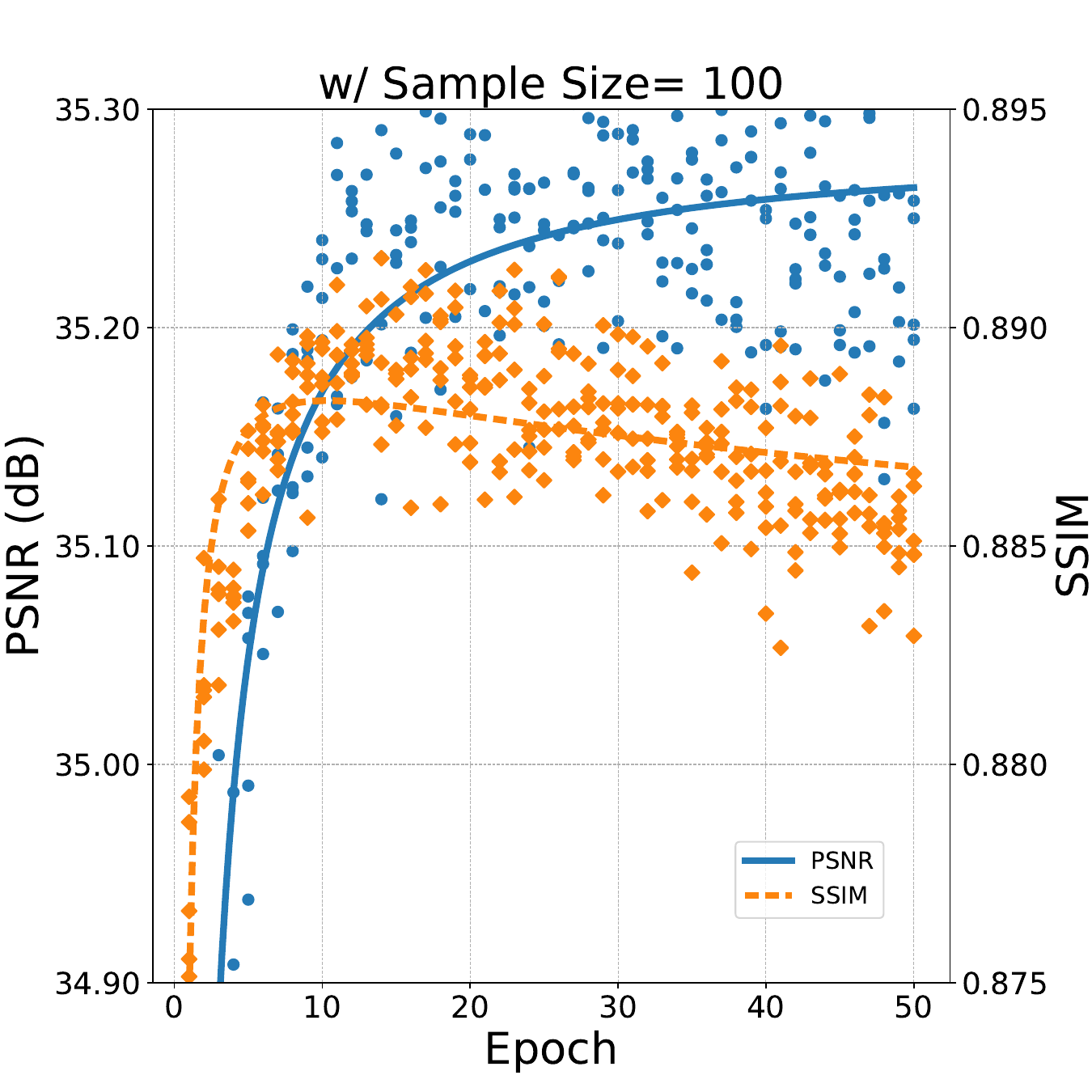}
    \end{subfigure}
    \hfill
    \begin{subfigure}{0.31\textwidth}
        \includegraphics[width=\linewidth]{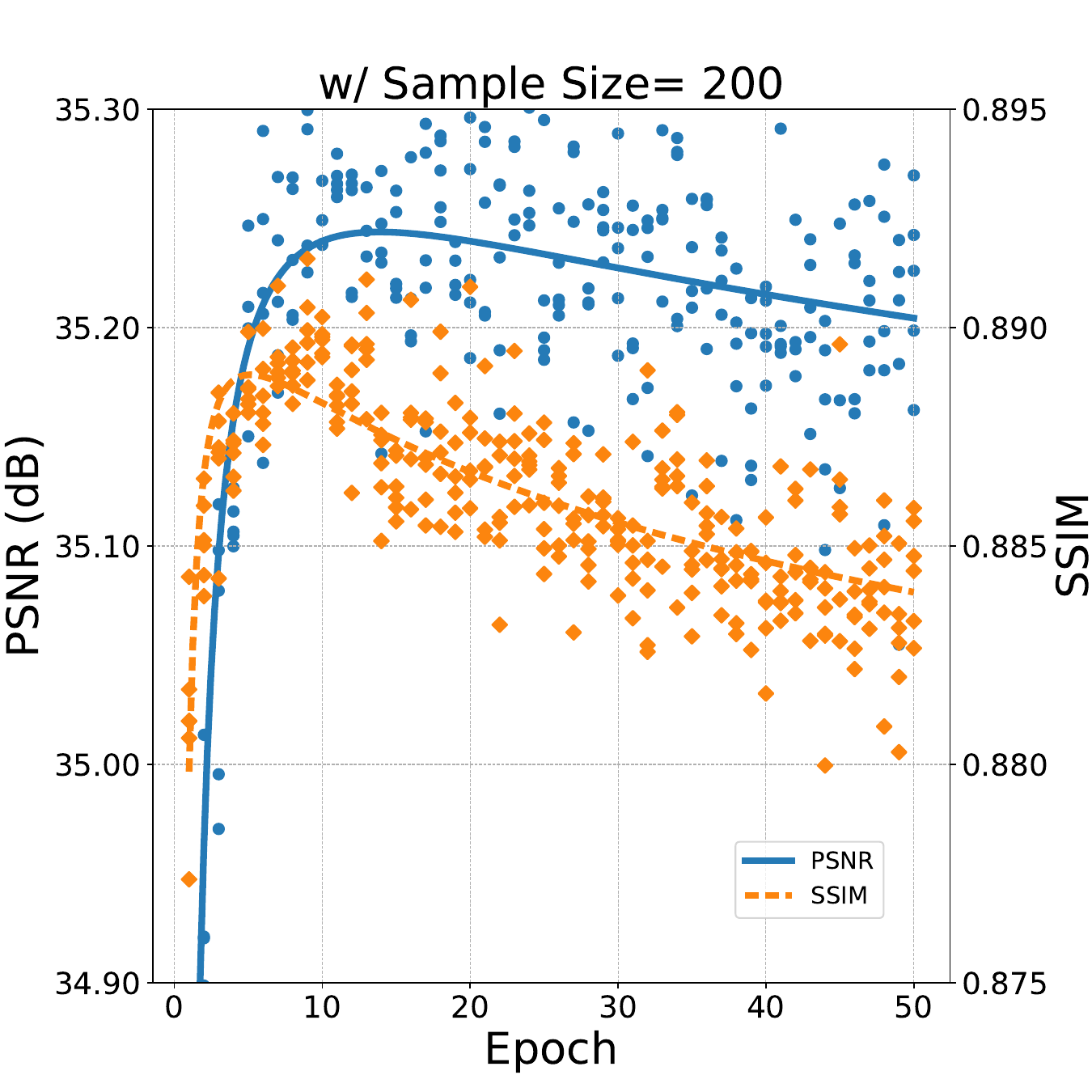}
    \end{subfigure}
    \hfill
    \begin{subfigure}{0.31\textwidth}
        \includegraphics[width=\linewidth]{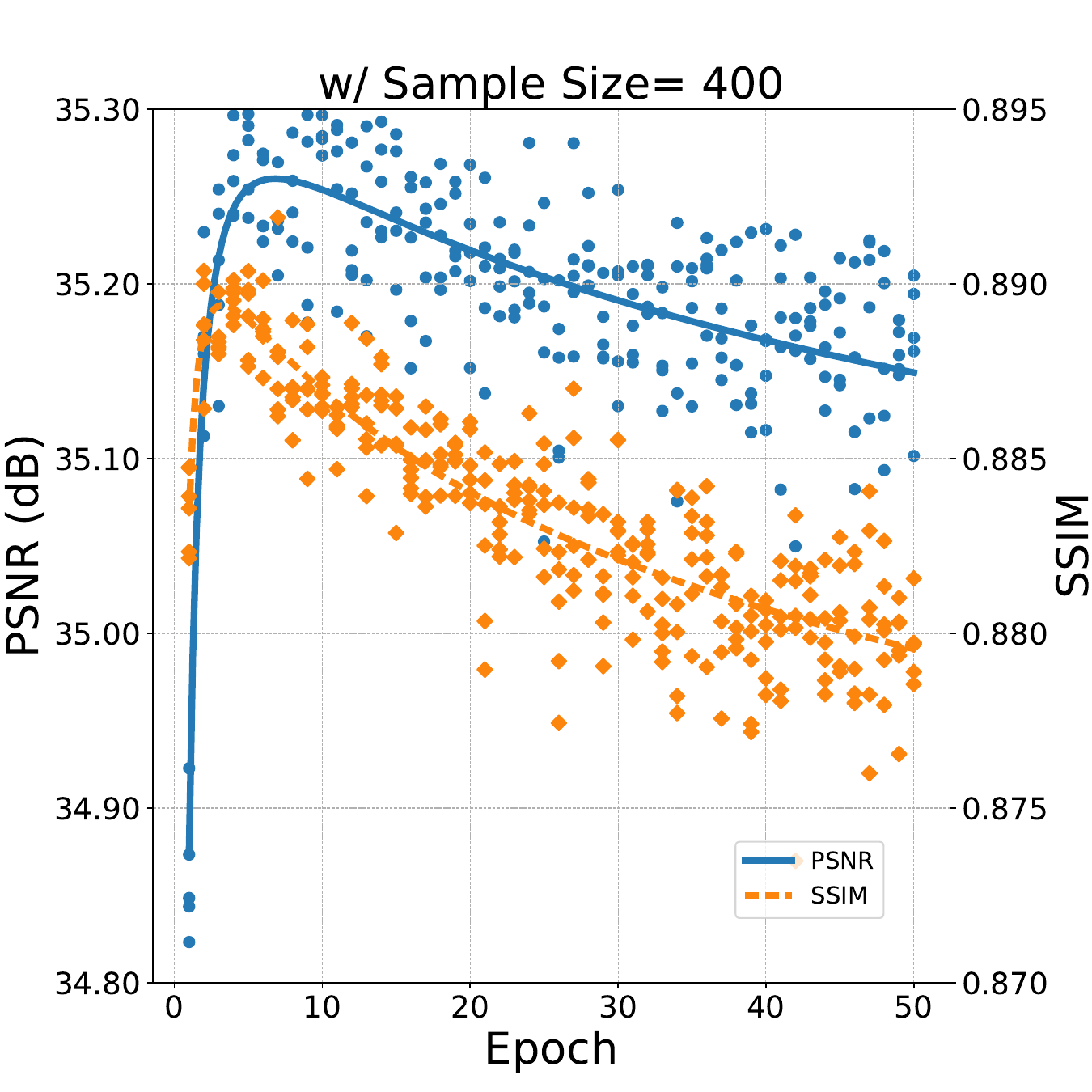}
    \end{subfigure}
    \hfill
    \begin{subfigure}{0.31\textwidth}
        \includegraphics[width=\linewidth]{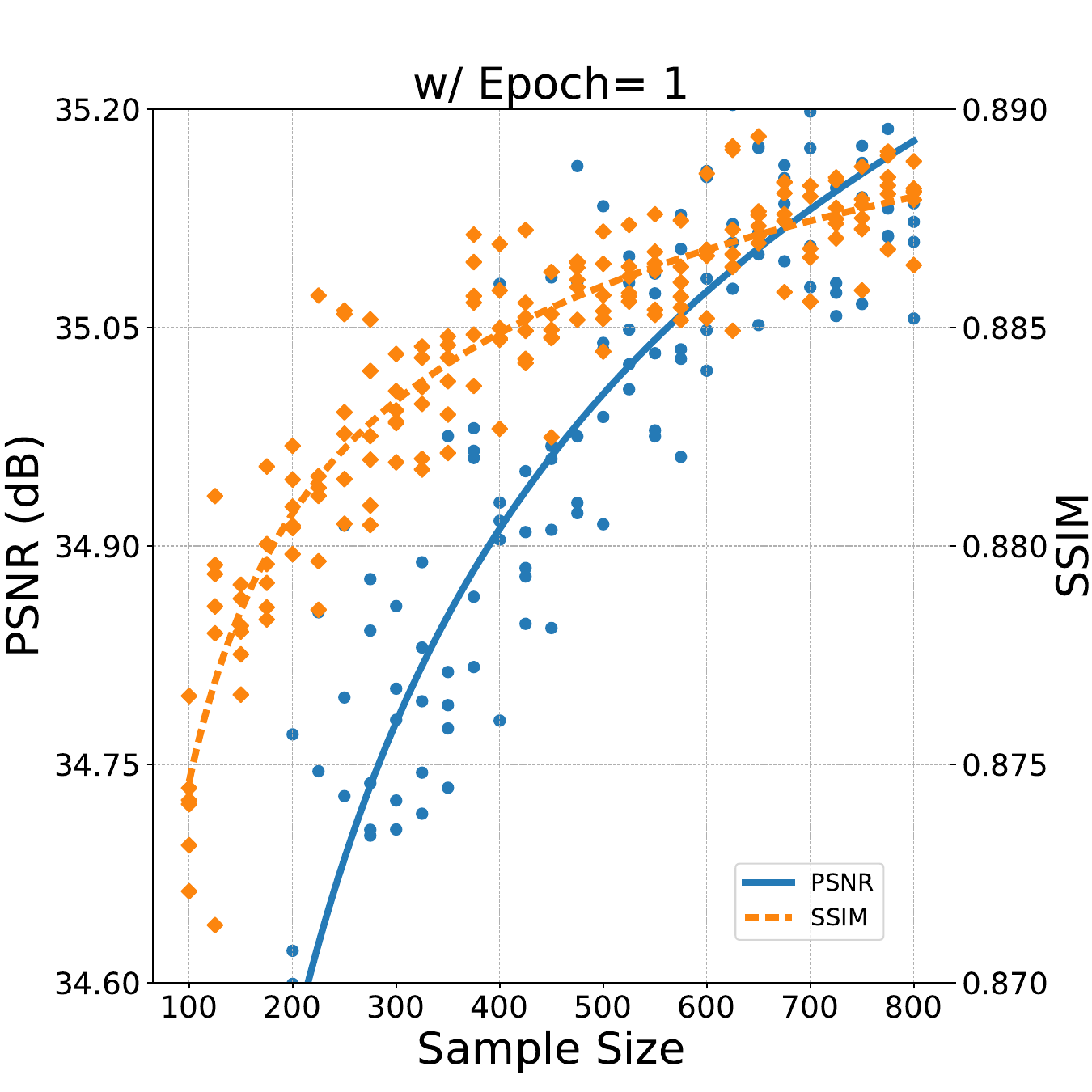}
    \end{subfigure}
    \hfill
    \begin{subfigure}{0.31\textwidth}
        \includegraphics[width=\linewidth]{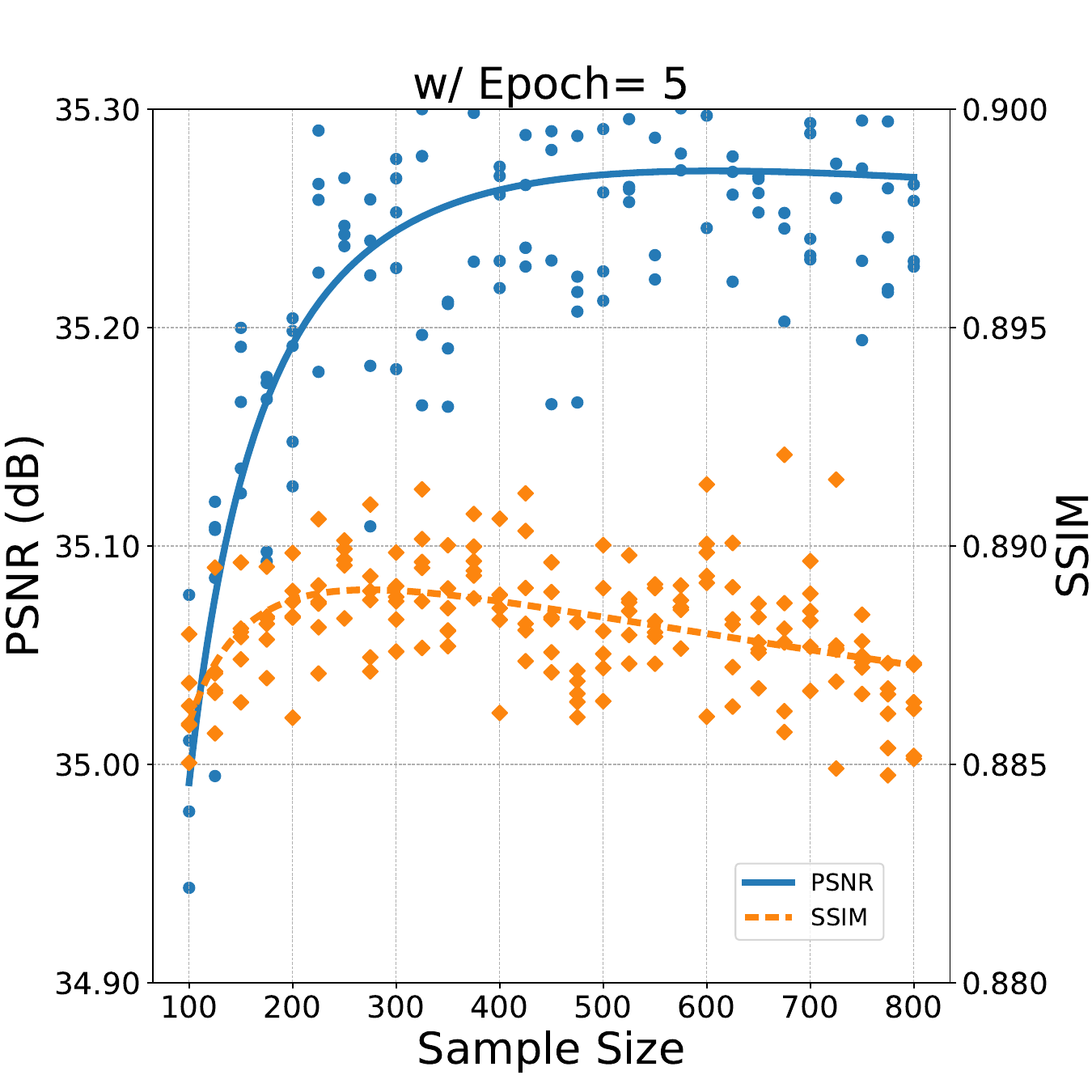}
    \end{subfigure}
    \hfill
    \begin{subfigure}{0.31\textwidth}
        \includegraphics[width=\linewidth]{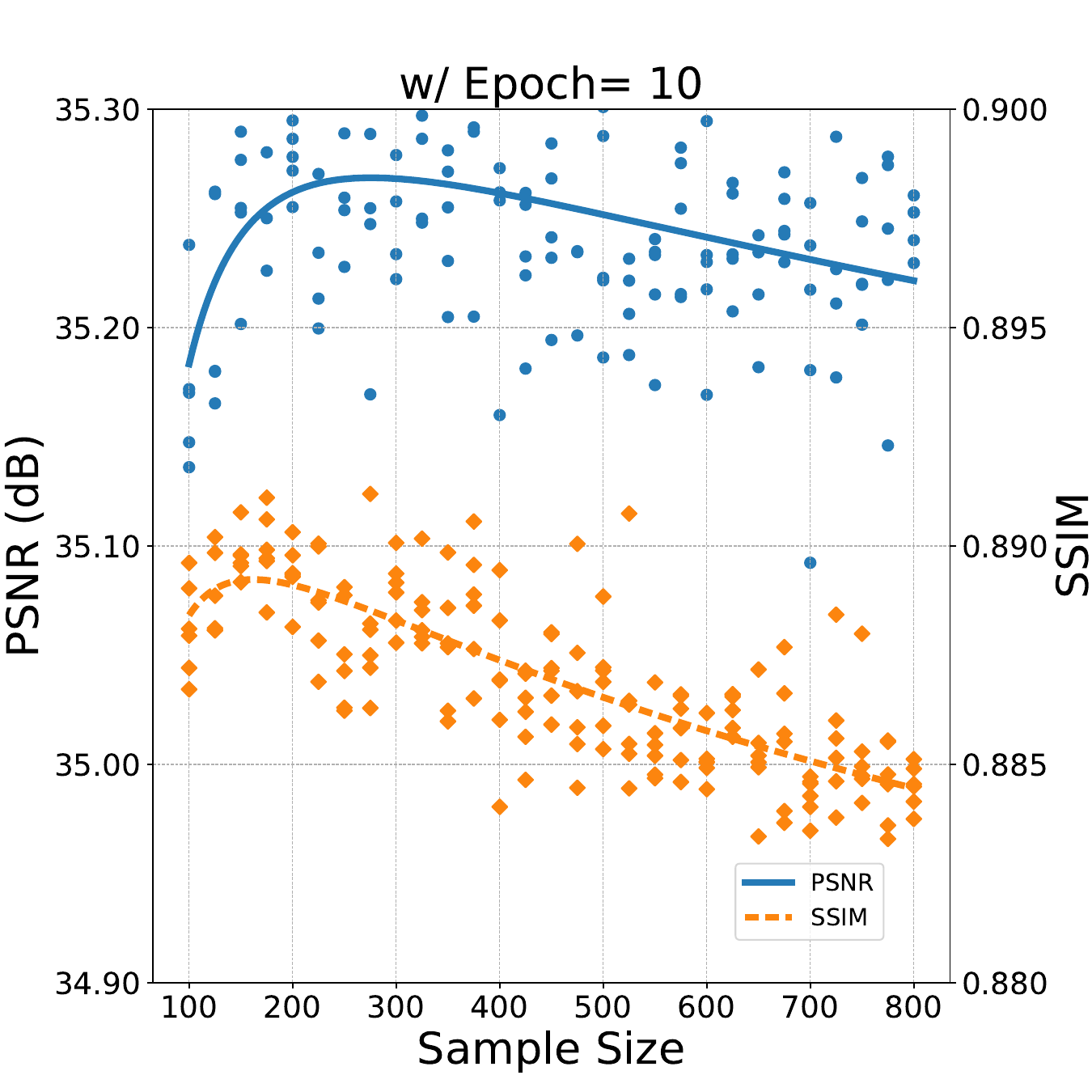}
    \end{subfigure}
    \caption{Denoising performance on a randomly selected subset, with the first row representing varying sample sizes (100, 200, 400) and the second row representing different epochs (1, 5, 10).}
    \label{fig:comparison}
\end{figure*}

We conduct a series of studies on the influences of Sample Size and Epoch as shown in Fig. \ref{fig:comparison}. The denoising performance boosts with increasing Epoch initially and decreases slowly after an optimal point under all three Sample Size settings, which is attributed to the easy overfitting because the model can only access to one noisy image for training. In contrast, with an appropriate Epoch setting, the denoising performance remains relatively steady, confirming that our Noise Injection module mitigates the overfitting and generates highly diverse training samples.

\subsection{Network Size}
FM2S utilizes a DNN with significantly fewer network parameters and achieves satisfying results. Here, we discuss network size to underscore the potential of lightweight networks in denoising. As Table \ref{table:networkSize} shows, adopting a large UNet architecture does not lead to better performance while leading to decreased efficiency. The phenomenon might be attributed to overfitting in single-image denoising. Our study reveals that a lightweight network with 3.5k parameters is enough to obtain good denoising effects while maintaining acceptable time consumption. This finding underscores the feasibility of applying deep-learning-based denoising methods in resource-constrained scenarios.

\begin{table}[htbp]
\centering
\footnotesize
\begin{tabular}{lcccc>{\columncolor{gray!20}}ccc} \toprule
Metric & 7.8M & 1.9M & 13.9k & 7.6k & 3.5k & 1.2k & 0.3k \\ \midrule
PSNR & 31.8 & 32.13 & 33.65 & \uline{34.09} & \textbf{34.15} & 33.84 & 33.28 \\
SSIM & 0.804 & 0.818 & 0.873 & \uline{0.884} & \textbf{0.886} & 0.882 & 0.867 \\
Time & 91sec & 61sec & 11sec & 8sec & 6sec & \uline{5sec} & \textbf{2sec} \\ \bottomrule
\end{tabular}
\caption{Performance and efficiency \textit{v.s.} network size. The data are obtained with the FMD dataset.}
\label{table:networkSize}
\end{table}

\section{Details and Hyperparameters}
The training details are presented in Table \ref{table:training_details}, and the hyperparameters for the SRDTrans and the FMD dataset are in Table \ref{table:Hyperparameters4SRDTrans} and Table \ref{table:Hyperparameters4FMD}, respectively.

\begin{table}[htbp]
    \centering
    \begin{tabular}{ccc}
    \toprule
         & Stage1 & Stage2 \\ \midrule
        Batch size & 1 & 1 \\
        Learning rate & 1e-3 & 1e-3 \\
        Adam $\beta$ & (0.9, 0.999) & (0.9, 0.999) \\
        Steps & 5 & $Epoch\times SampleSize$ \\
    \bottomrule
    \end{tabular}
    \caption{Training details.}
    \label{table:training_details}
\end{table}

\begin{table}[htbp]
\centering
\begin{tabular}{clc}
\toprule
Dataset & HypParam & Value \\ \midrule
\multirow{6}{*}{SRDTrans} & $\lambda$ & 5 \\ 
 & stride & 5 \\
 & filter & 3 \\
 & $k_g$ & 60 \\
 & $k_p$ & 30 \\
 & $\lambda_p$ & 150 \\ \bottomrule
\end{tabular}
\caption{Hyperparameters for the SRDTrans dataset \cite{li2023spatial}.}
\label{table:Hyperparameters4SRDTrans}
\end{table}

\begin{table}[htbp]
\centering
\footnotesize
\begin{tabular}{clccccc}
\toprule
Type & HypParam & avg1 & avg2 & avg4 & avg8 & avg16 \\ \midrule
\multirow{6}{*}{Confocal} & $\lambda$ & \multicolumn{5}{c}{2} \\
 & stride & \multicolumn{5}{c}{75} \\
 & filter & \multicolumn{5}{c}{3} \\
 & $k_g$ & 200 & 125 & 70 & 10 & 5 \\
 & $k_p$ & 30 & 95 & 195 & 240 & 650 \\
 & $\lambda_p$ & 70 & 285 & 485 & 650 & 1400 \\ \midrule
\multirow{6}{*}{TwoPhoton} & $\lambda$ & \multicolumn{5}{c}{2} \\
 & stride & \multicolumn{5}{c}{75} \\
 & filter & \multicolumn{5}{c}{3} \\
 & $k_g$ & 175 & 150 & 90 & 20 & 15 \\
 & $k_p$ & 30 & 85 & 300 & 185 & 850 \\
 & $\lambda_p$ & 60 & 300 & 480 & 600 & 3800 \\ \midrule
\multirow{6}{*}{WideField} & $\lambda$ & \multicolumn{5}{c}{1} \\
 & stride & \multicolumn{5}{c}{75} \\
 & filter & \multicolumn{5}{c}{11} \\
 & $k_g$ & 220 & 220 & 60 & 20 & 1 \\
 & $k_p$ & 45 & 100 & 650 & 600 & 1500 \\
 & $\lambda_p$ & 2000 & 2500 & 3500 & 4000 & 4800 \\ \bottomrule
\end{tabular}
\caption{Hyperparameters for the FMD dataset \cite{zhang2019poisson}.}
\label{table:Hyperparameters4FMD}
\end{table}

\end{document}